\documentclass[aps,pra,groupedaddress,reprint,floatfix,notitlepage,nobalancelastpage]{revtex4-2}
\usepackage[T1]{fontenc}       
\usepackage{microtype}         
\usepackage{graphicx,amsmath,amssymb,bm,color,engord,xfrac,braket}\usepackage{bbold}
\usepackage[usenames,dvipsnames]{xcolor}
\usepackage[colorlinks,colorlinks,urlcolor=Magenta,citecolor=Magenta,linkcolor=blue]{hyperref}
\usepackage{newtxtext}
\usepackage{newtxmath}
\newcommand{\br}{{\bm r}}

\newcommand{\bone}{\mathbf{1}}
\newcommand{\curP}{{\cal P}}
\newcommand{\kf}{k_{\rm F}}
\newcommand{\vf}{v_{\rm F}}

\newcommand{\bk}{{\bm k}}
\newcommand{\kb}{k_{\rm B}}
\newcommand{\bq}{{\bm q}}\newcommand{\xh}{\hat{x}}
\newcommand{\nf}{n_{\rm F}}
\newcommand{\as}{a_s}
\newcommand{\tc}{{T_{\rm c}}}
\newcommand{\grad}{{\bm{\nabla}}}
\newcommand{\curK}{{\cal K}}
\newcommand{\curR}{{\cal R}}
\newcommand{\curE}{{\cal E}}
\newcommand{\Deltab}{\bar{\Delta}}
\newcommand{\be}{\begin{equation}}
\newcommand{\ee}{\end{equation}}
\newcommand{\bea}{\begin{eqnarray}}
\newcommand{\eea}{\end{eqnarray}}
\newcommand{\bse}{\begin{subequations}}
\newcommand{\ese}{\end{subequations}}

\newcommand{\ph}{\hat{\bf p}}
\begin{document}

\title{Pairing near the boundary of a box-shaped trap in the BEC-BCS crossover}
\author{Kelly R. Patton}
\email[\hspace{-1.4mm}]{kellypatton@uncc.edu}
\affiliation{Department of Physics \& Optical Science, University of North Carolina at Charlotte, Charlotte, NC, 28223, USA}
\author{Daniel E. Sheehy}
\email[\hspace{-1.4mm}]{sheehy@lsu.edu}
\affiliation{Department of Physics \& Astronomy, Louisiana State University, Baton Rouge, Louisiana 70803, USA}
\date{July 31, 2026}
\begin{abstract}
We study pairing of attractively interacting fermions confined to a box-shaped trap. In contrast to the infinite translationally-invariant case, where the local pairing order is  spatially uniform and undergoes the Bose-Einstein Condensate to Bardeen-Cooper-Schrieffer (BEC-BCS) crossover as interactions are varied, in this case the local pairing is expected to vary rapidly near the edge of the box. We address this problem in the limit of a semi-infinite superfluid, finding that the nature of the edge pairing depends sensitively on the coupling.   
The local pairing exhibits Friedel-like oscillations in the weak coupling BCS regime that are suppressed with increasing coupling strength towards the BEC regime.  
\end{abstract}
\maketitle
\section{Introduction}

A long-standing problem in the related fields  of 
superconductivity  in electronic materials and superfluidity  of cold fermionic atomic gases concerns how the system boundary affects Cooper pairing and other superfluid properties.  Some of the questions of interest
include  how the local pairing amplitude $\Delta(\br)$ varies
near the edge of a superconductor (or near an interface with another material)~\cite{deGennes,StojkovicPRB1993,StojkovicPRB1994}, how the local pairing or superconducting transition temperature
is affected by confinement or finite-size effects~\cite{BlattThompson,GarciaGarcia2008,ValentinisPRB16,Valentinis2016,ValentinisBerthod,Hainzl2022,Roos2025}, the possibility of Andreev bound states at the system edge~\cite{Sauls2021}, and 
whether pairing may be enhanced at the system boundary, leading 
to surface superconductivity~\cite{Giamarchi1990,DorseyPRB1995,SamoilenkaBabaev2020,Croitoru2020,Barkman2019}.

Our interest in this subject was spurred by the development of spatially homogeneous “box”-shaped traps for confining cold bosonic and fermionic atomic gases~\cite{Gaunt,Mukherjee,Lopes2018,Eigen2017,Hueck2018,Garratt2018,Fletcher2018,Yan2018,Baird2019,Patel,NavonNaturePhysics2021}. While homogeneous in the bulk, such box traps exhibit a rapidly varying single-particle potential near the box edge. A natural question arises: How is the local pairing amplitude  $\Delta(\br)$ modified by such an edge? Although our study is motivated by the cold-atom setting of a paired-fermion superfluid (SF), such a box trapping potential can also approximately describe the edge of an electronic superconductor (SC), so that our study of this question applies to both settings.  

In addressing this question, one theoretical issue concerns
the fact that for short-ranged interactions the self-consistent gap equation  for 
$\Delta(\br)$ is {\em cut-off dependent\/} within the 
Bardeen-Cooper-Schrieffer (BCS) theory that applies to the 
Bose-Einstein Condensate (BEC)-BCS crossover~\cite{Gurarie}. The gap
equation 
involves an integral (over energies or momenta) that
must be cut off at an ultraviolet (UV) scale.  In the case of
conventional electronic superconductors, this UV scale is typically
the Debye frequency.  In the case of cold-atom fermionic
superfluids, this UV physics can be handled by using 
the Lippmann-Schwinger equation, which relates the bare
inter-fermion coupling $\lambda$ to the $s$-wave scattering length $a_{s}$:
\begin{equation} 
   \label{lambdaasrelation}
\frac{1}{\lambda} = \frac{m}{4\pi \as\hbar^2} - \int_\Lambda \frac{{\rm d}^3 k}{(2\pi)^3} \frac{1}{2\epsilon_\bk},  
 \end{equation}
 where $\epsilon_\bk = (\hbar \bk)^2/(2m)$ is the non-interacting single-particle energy and $\Lambda$ is the UV cutoff.  The idea
 is that although the gap equation is also dependent on $\Lambda$,
 when one combines it with Eq.~(\ref{lambdaasrelation}), the
 resulting \lq\lq renormalized\rq\rq\ gap
 equation is insensitive to UV physics, i.e.,
 one can take the limit of $\Lambda \to \infty$ and get 
 universal results.  Assuming translational invariance (in the bulk of an SF or SC),  
 this procedure is straightforward (since $\Delta(\br)=
 \Delta$ can be
 assumed spatially uniform), leading to a simple formula for $\Delta$ at a given $\as$ (effectively a renormalized coupling)  and temperature $T$.  

 \begin{figure}[t]
  \centering
  \includegraphics[width=\columnwidth]{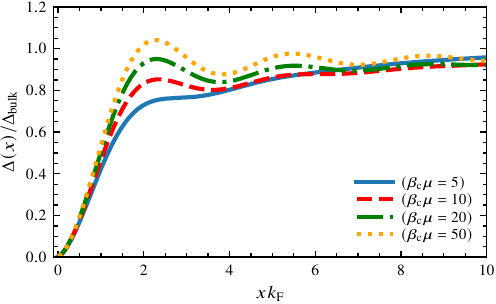}
  \caption{(Color online) Plot of the local pairing
 amplitude $\Delta(x)$ vs. position $x$ in the regime $T<T_{\rm c }$
 for a semi-infinite "half slab" Fermi superfluid, normalized
 to the bulk value $\Delta_{\rm bulk}$.  The distinct
 curves show systems at different values of the coupling-dependent transition temperature $\tc$ 
 (with the figure legend showing $\beta_{\rm c}\mu\equiv \mu/(\kb T_{\rm c})$, where $\mu$ is the chemical potential). 
 The system temperature $T$ for each curve is slightly below 
 $\tc$, with   $\beta\mu = 1.1 \beta_\mathrm{c}\mu$ or $(T_{\rm c}-T)/T_{\rm c}\approx 0.09$. As one can see, the pairing rises rapidly  from zero at the boundary to approach the bulk value on a scale of roughly $(2\kf)^{-1}$. The local pairing also exhibits Friedel-like oscillations near the boundary, which also have a wavelength of roughly  $(2\kf)^{-1}$.    These oscillations become much more prominent as $T_{\rm c}$ is lowered, i.e., in the weak-coupling BCS regime. This reflects the fact that at higher $T_{\rm c}$ a sharp Fermi surface is no longer present.   
 }
  \label{fig:Various Temperatues}
\end{figure}

 In studying boundary or edge effects in an SC or SF,
 we need to be careful about the cutoff/regularization issue for two  reasons: first,
 the proper regularization of the gap equation  is not as simple
 in the case where $\Delta(\br)$ is not spatially uniform.  Secondly,
 in the case of an SF in a box-shaped trap, we expect the strongest 
 spatial variation of $\Delta(\br)$ to be near the edge of the box.  Since spatial variations on short length scales correspond (in  Fourier space) to large wavevectors, it is essential to have a formalism
 in which any cutoff dependence is consistently
 handled. 
Said more succinctly, we need to ensure that any predicted edge effects are independent of the choice of the cutoff scale and
   universal in the limit of $\Lambda \to \infty$.  We note that
 an alternate approach to regularizing
 the BCS problem involves using the pseudopotential prescription, as studied in the cold-atom context by Bruun et al.~\cite{BruunEPJD1998} and a similar approach was applied to  nuclear systems in Ref.~\cite{BulgacPRL2002}.

As stated above, our main interest concerns 
how the local pairing $\Delta(\br)$ varies near a sharp boundary, such as the edge of a box-shaped trap for confined fermionic atomic gases or a similar boundary in an electronic superconductor.  We note that
if the external trapping potential $V(\br)$ is slowly varying then several methods
exist to find the local superfluid pairing.  For example,  Ginzburg-Landau 
theory~\cite{Gorkov1959P}, which amounts to a long wavelength approximation for $\Delta(\br)$,  holds for temperatures near the transition temperature $\tc$.    Within BCS theory itself, one can also use a semi-classical local density approximation (LDA) to account for the effects of an inhomogeneous system.  This approximation amounts to replacing the bulk chemical potential $\mu$ with a spatially varying one given by $\mu(\br ) = \mu - V(\br)$.  Extending  LDA to include gradient terms  is possible but challenging \cite{SchuckEPJA2023}.  
Other coarse-graining methods of the Bogoliubov-de Gennes equations have been
developed in Refs.~\cite{Simonucci2014,Simonucci2017}.
We note that for box-like trapping potentials GL becomes unreliable \cite{StojkovicPRB1993}  and  LDA and its extensions break down, since $V(\br)$ is no longer slowly varying.  To avoid
such issues, in the following, we analyze the pairing instability of a confined SC/SF without making such semiclassical or long-wavelength approximations.

\subsection{Outline and Main Results}
We now present the outline of the remainder of the paper and summarize our main results.   %
In Sec.~\ref{sec:modelham} we introduce a general standard model
Hamiltonian for attractively interacting fermions that can 
describe the well-known BCS-BEC crossover as a function of the coupling.
Our model includes a UV cutoff $\Lambda$; 
nonetheless, we explicitly show how our results are independent of this 
choice.  
To study edge effects on pairing, in Sec.~\ref{sec:hsfs} we specialize to a specific geometry, namely, a semi-infinite
3D system defined for all $y$ and $z$ and for $x>0$. This amounts to  imposing  hard-wall boundary conditions at $x=0$ for all  $y$ and $z$, i.e., $\Delta(x=0,y,z)=0$. Such an infinitely steep barrier is representative of the box-like potentials produced in ultra-cold atomic gas experiments  or a superconductor-insulator interface.

In Sec.~\ref{sec:pairingkernelEigenfunctions}, we describe
the superconducting instability at a critical temperature $\tc$ 
as an integral equation with
a cutoff-dependent kernel $\curK^\Lambda(x,x')$ and study
the eigenfunctions of  the kernel.   As
in standard quantum mechanics, one may expect either 
continuum (delta-normalizable) or 
bound-state (normalizable) eigenfunctions 
(or both).  We numerically find only continuum eigenfunctions in our present model,
with no evidence for discrete bound states 
(indicating the absence of \lq\lq surface superconductivity\rq\rq\ ).  These continuum eigenfunctions exhibit
oscillatory behavior of the form $\cos(qx)$
at $x\to \infty$ (with $q>0$ being the eigenfunction quantum number), while also vanishing for $x\to 0$ 
(consistent with the boundary condition in our model).
In the deep BCS regime, this vanishing of the
eigenfunctions takes the form of a sudden drop 
near $x\to 0$ accompanied by rapid ``Friedel'' 
oscillations at a scale $2\kf$ with $\kf$
the Fermi wavevector.  We find that these Friedel oscillations are suppressed with increasing coupling strength, moving towards the BEC regime.  
In Sec.~\ref{sec:eeignorm}, we describe the proper normalization of these
continuum eigenfunctions of $\curK^\Lambda(x,x')$.
Although the eigenvalues are cutoff dependent
(and must be renormalized using Eq.~(\ref{lambdaasrelation})), the eigenfunctions 
are cutoff independent (i.e., universal
at $\Lambda\to \infty$), allowing us to 
construct a renormalized cutoff-independent
pairing kernel $\curK^R(x,x')$.

In Sec.~\ref{sec:edgepairing}, we 
use our result for $\curK^R(x,x')$ to
study pairing slightly
below $\tc$ by adding a Ginzburg-Landau-like
nonlinear term to our model. This nonlinear term is used to establish bulk pairing for $x\to \infty$.  We find that 
spatial variation of pairing near the boundary of a semi-infinite
superfluid can arise from two sources: 1) Short-distance edge effects inherited from the pairing kernel eigenfunctions and 2) A long distance spatial
variation of pairing controlled by the 
coherence length $\xi\propto 1/\sqrt{\tc-T}$. In
Fig.~\ref{fig:Various Temperatues}, we show results
for the edge pairing at $T<\tc$, normalized to the bulk
value, for various coupling
values. All four curves are at the same normalized temperature
$T/\tc\simeq 0.91$, but for different 
coupling values across the BEC-BCS crossover
showing how the edge pairing changes.  
The couplings can be expressed in terms of 
$\kf \as$  or in terms of $\kb \tc/\mu$ (with
$\kb$ the Boltzmann constant and $\mu$ the chemical potential), with $\tc$ increasing monotonically with increasing coupling strength as we move from the BCS to the BEC regime.   
Starting in the deep BCS regime (yellow dotted curve at 
$\kb \tc/\mu \simeq 0.02$, or $\kf \as  \simeq -0.46$), the edge pairing shows strong Friedel oscillations that are inherited from
similar behavior in the kernel eigenfunctions.  With increasing
coupling these oscillations are suppressed and 
replaced with a smooth increase towards the bulk, as seen in the green dot-dashed ($\kb \tc/\mu \simeq 0.05$ or $\kf \as  \simeq -0.63$), red dashed ($\kb \tc/\mu \simeq 0.1$ or $\kf \as  \simeq -0.86$) and  blue solid ($\kb \tc/\mu \simeq 0.1$  or $\kf \as  \simeq -1.39$)
curves. 

In Sec.~\ref{sec:concl} we provide
some concluding remarks.
Appendix~\ref{app:dvpi} discusses how the integral of a product of principal-value distributions leads to a delta distribution.
Appendix~\ref{app:GL} reviews relevant
aspects of Ginzburg-Landau theory.
Appendix~\ref{app:bound} derives a bound
on the spectrum of the pairing kernel $\curK^\Lambda$, and 
Appendix~\ref{app:numdet} presents
details of our numerical method.

\section{Model Hamiltonian}
\label{sec:modelham}
In this section we start from the following generic model
Hamiltonian for an untrapped 
spin-$\sfrac{1}{2}$ Fermi gas with contact interactions:
\begin{align}  
H = \int {\rm d}^3 r\, &\Bigg[\sum_{\sigma = \uparrow,\downarrow}\Psi_\sigma^\dagger(\br)
\Big(\frac{\ph^2}{2m} -\mu\Big) \Psi_\sigma(\br) \nonumber
\\ & 
+ \lambda\,\Psi^{\dagger}_\uparrow(\br) 
\Psi^{\dagger}_\downarrow (\br) \Psi^{}_\downarrow (\br) \Psi^{}_\uparrow (\br) \Bigg],
\label{Eq:originalHam}
\end{align}
where $\ph = -i\hbar \grad$ is the momentum operator with $\hbar$ the reduced Planck's constant,
$m$ the fermion mass, $\mu$ the chemical potential, and
$\lambda$ the bare delta function interaction coupling parameter.
Here, the $\Psi_\sigma(\br)$ are the usual
fermion field operators obeying the 
anti-commutator relation
\be
\{ \Psi_\sigma(\br), \Psi_{\sigma'}^\dagger(\br')\} = \delta_{\sigma,\sigma'}
\delta(\br-\br').  
\ee
As of yet, we have not yet specified the domain of our system. Below we will focus
on the case of a semi-infinite system, with a hard-wall boundary at $x=0$ as a model
for the vicinity of the boundary in a box-shaped trap.  Before turning to this, we briefly recall
the conventional BEC-BCS mean-field theory for a uniform (i.e. infinite) system.
There, one finds a mean-field solution for the spatially-uniform pairing amplitude 
$\Delta = \lambda\langle \Psi_\downarrow(\br) \Psi_\uparrow(\br)\rangle$
that satisfies the gap equation: 
\be
\frac{1}{\lambda} = - \int\limits_{k<\Lambda} \frac{{\rm d}^3 k}{(2\pi)^3} \frac{\tanh \frac{E_\bk}{2T}}{2E_\bk},
\label{Eq:gapcutoff}
\ee
where $E_\bk = \sqrt{\xi_\bk^2 +|\Delta|^2}$ with $\xi_\bk = \epsilon_\bk -\mu$.

As is well-known, the integral on the right side of 
Eq.~(\ref{Eq:gapcutoff}) must be regularized, which can be done, for example, by introducing a 
ultraviolet (UV) cutoff $\Lambda$ as we have done here.  Although this seems to imply
that observable quantities will be cutoff-dependent, the 
standard picture is that the bare coupling parameter $\lambda$
is itself related to the $s$-wave scattering length $\as$ in a
cutoff-dependent way via the Lippmann-Schwinger equation,
Eq.~(\ref{lambdaasrelation}) above. 
    We note that in the following we also define the renormalized coupling 
     $g\equiv 4\pi \as\hbar^2/m$.
Upon plugging  Eq.~(\ref{lambdaasrelation}) into 
Eq.~(\ref{Eq:gapcutoff}), we can combine the integrals and
take the limit  $\Lambda\to \infty$;
\be
  \frac{m}{4\pi \as\hbar^2}  = \frac{1}{g} = 
  - \int \frac{{\rm d}^3 k}{(2\pi)^3} \left(\frac{\tanh \frac{E_\bk}{2T}}{2E_\bk} - \frac{1}{2\epsilon_\bk}\right),
  \label{eq:combinedintegral}
\ee
which yields a cutoff-independent result for the pairing amplitude at a
given temperature $T$ and scattering length
$\as$.  

Note that the preceding steps, which regularize the theory, are simplest when
we can assume a spatially uniform pairing amplitude
$\Delta$.  In the next section we introduce a
model for a ``half-slab'' Fermi superfluid in which
$\Delta$ is expected to vary spatially, with a key
question being how to consistently implement
the regularization in this case.

\section{Half-slab Fermi superfluid}
\label{sec:hsfs}
To study pairing effects near the boundary of a 
 superconductor (SC) or atomic-fermion superfluid (SF),
we consider a semi-infinite half-slab system that
is infinite along the $y$ and $z$ directions (or a large finite
system, with periodic boundary conditions, in the thermodynamic limit).  The system is semi-infinite along the $x$ direction, with the system defined
for $x>0$ and with a hard-wall boundary condition at $x=0$.  We expect that the half-slab system is a good approximation sufficiently close to the walls of a SF or SC.  

The first step is to define a convenient basis for the single-particle states.  Translational invariance along the $y$ and $z$ directions implies that pairing is independent of these coordinates, allowing us to Fourier transform with 
respect to the $y$ and $z$ directions.  Along the $x$
direction, we'll use a few different basis sets, 
two of which are the sine and cosine basis sets
\bse
\label{Eq:cossinbasis}
\bea
\psi_k(x) &=& \sqrt{2}\sin(kx), \\
\chi_Q(x) &=& \sqrt{2}\cos(Qx),
\label{eq:cosbasisfns}
\eea
\ese
both of which are orthonormal and complete (in
the distributional sense) in the regime $x>0$ 
(see Eqs.~(\ref{eqs:kkqq}) and (\ref{eqs:kkqqcomp})).

Following standard BCS/Bogoliubov theory, we find for the pairing
instability at $\tc$, the linearized integral equation:
\be
\label{Tc eigenvalue equation in real space}
\frac{1}{\lambda}\Delta(x) = \int_0^\infty {\rm d}x'\, \curK^\Lambda(x,x') \Delta(x'),
\ee
where the pairing kernel $\curK^\Lambda(x,x')$ is given by 
\bea
&&\curK^\Lambda(x,x') = \int_0^\infty \frac{{\rm d}k_1}{\pi}  \int_0^\infty \frac{{\rm d}k_2}{\pi}\,
f^\Lambda(k_1,k_2)  
\\
&&\qquad\qquad \times 
\psi_{k_1}(x) \psi_{k_1}(x') \psi_{k_2}(x) \psi_{k_2}(x') ,
\nonumber 
\eea
where the superscript $\Lambda$ denotes that the kernel
is explicitly dependent on the UV cutoff $\Lambda$ via the integrand
\bea
\label{eq:integrandF}
f^\Lambda(k_1,k_2) = \int\limits^{}_{k_\parallel<\Lambda}
\frac{{\rm d}^2 k_\parallel}{(2\pi)^2}
 \frac{\nf(\xi_{k_{\parallel} k_1}) - \nf(-\xi_{k_{\parallel} k_2})}
     {\xi_{k_{\parallel} k_1}+\xi_{k_{\parallel} k_2}},
\eea
where $\nf(x) = \frac{1}{{\rm e}^{x/T}+1}$ is the Fermi 
distribution at temperature $T$, $\xi_{k_{\parallel} k} = 
\frac{1}{2m}(k_{\parallel}^2 + k^2) -\mu$ is the energy with
$\bk_\parallel$ the wavevector parallel to the slab (i.e. 
along $y$ and $z$) and $k$ the perpendicular momentum.  Note that here and below we 
take $\kb = \hbar = 1$.  

\section{Eigenfunctions of the Pairing Kernel}
\label{sec:pairingkernelEigenfunctions}
To analyze the pairing instability of the half-slab 
Fermi superfluid, we seek the eigenfunctions of 
 $\curK^\Lambda(x,x')$. To find their properties, 
 we first perform a double
cosine Fourier transform with respect to the basis
functions in Eq.~(\ref{eq:cosbasisfns}), defining 
\bea
\nonumber 
\langle Q|\curK^\Lambda|Q'\rangle &\equiv &
\int_0^\infty {\rm d}x \int_0^\infty {\rm d}x' \,
\chi_Q(x) \curK^\Lambda(x,x') \chi_{Q'}(x')
\\
& = & \pi \delta(Q-Q')
\curE^\Lambda(Q) 
- \frac{1}{2} f_{Q,Q'}^\Lambda,
\label{Eq:kappamatrix}
\eea
where we defined 
\bea
f_{Q,Q'}^\Lambda
&\equiv & f^\Lambda(\frac{1}{2}(Q-Q'),\frac{1}{2}(Q+Q')),
\label{eq:effQQ}
  \\
\label{eq:epsilonLambdaQ}
\curE^\Lambda(Q)&\equiv & 
\frac{1}{2}\int_0^\infty \frac{{\rm d}Q'}{\pi} f_{Q,Q'}^\Lambda .
\eea
Using the preceding definitions, it is straightforward
to check that $\int_0^\infty \frac{{\rm d}Q}{\pi} \langle Q|
\curK^{\Lambda}|Q'\rangle = 0$, and similarly for the $Q'$ integral.
These amount, in real space, to the conditions $\curK^\Lambda(0,x') = 0$
and $\curK^\Lambda(x,0) = 0$, respectively.

\begin{figure}[t]
  \centering
  \includegraphics[width=\columnwidth]{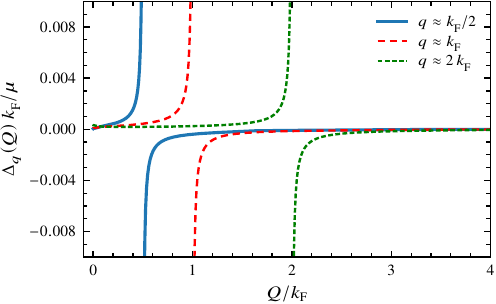}
  \caption{(Color online) Shown here are representative eigenstates of the pairing kernel $\curK^\Lambda$ with the eigenvalue equation given in  Eq.~\eqref{eq:eigenvaluerelation}.   
  The plot shows three such eigenstates (with distinct wavevector quantum numbers)
  in the low-$\tc$ BCS regime of $\beta_\mathrm{c}\mu=100$
    as a function of wavevector $Q$ 
  (normalized to the Fermi wavevector) }
  \label{fig:LowTcEigenstates in Momentum space}
\end{figure}

\subsection{Numerical Analysis of Pairing Kernel}
We see that the kernel $\curK^\Lambda$ has a simple
structure in the cosine basis, consisting of a \lq\lq diagonal\rq\rq\
piece 
\be
\curK^\Lambda_{\rm diag.} 
= \pi \delta(Q-Q')\curE^{\Lambda}(Q),
\ee
plus an off-diagonal
part:
\be
\curK^\Lambda_{\rm off.}  = 
-\frac{1}{2}f_{Q,Q'}^\Lambda,
\ee
that is needed to enforce the boundary conditions.
In the absence of this off-diagonal
part, the spectrum of the kernel would be given by that of $\curK^\Lambda_{\rm diag.}$, i.e., equal to $\curE^\Lambda(q)$ (which we call the bulk eigenvalues),
with singular generalized eigenfunctions $\propto 
\delta(Q-q)$.  
How does $\curK^\Lambda_{\rm off.}$
modify the continuous (essential) spectrum of 
$\curK^\Lambda_{\rm diag.}$? In fact, a theorem
due to Weyl~\cite{Weyl} (also discussed in
Reed and Simon~\cite{ReedSimon} and 
employed recently in the context of Migdal-Eliashberg theory~\cite{Elezaby}) implies that since  $\curK^\Lambda_{\rm off.}$ is 
compact, it cannot
modify the essential spectrum of 
$\curK^\Lambda_{\rm diag.}$.  Compactness of 
the off-diagonal piece, which holds because
\be
\int_0^\infty {\rm d}Q \int_0^\infty {\rm d}Q' |f^{\Lambda}_{Q,Q'}|^2<\infty, 
\ee
therefore implies that the essential spectrum of the full kernel $\curK^\Lambda = \curK^\Lambda_{\rm diag.}+\curK^\Lambda_{\rm off.}$ is given by $\curE^\Lambda(q)$.

To confirm this expectation, and find the eigenfunctions and eigenvalues
of $\curK^\Lambda$, the simplest approach is to 
discretize the wavevectors $Q$ and $Q'$ on a large grid 
(in the presence of a chosen value of the UV cutoff),
evaluate Eq.~(\ref{Eq:kappamatrix}) on the grid points, and 
numerically find the eigenvectors and eigenvalues.  
Typical results for this are shown in 
Fig.~\ref{fig:LowTcEigenstates in Momentum space}, where we show
three eigenfunctions that show singular behavior as a function of
wavevector $Q$.  The location of the singularity, $q$, serves as a label 
(or quantum number) for
the eigenfunctions, which we write as $\Delta_{q}(Q)$.  
\begin{figure}[t]
  \centering
  \includegraphics[width=\columnwidth]{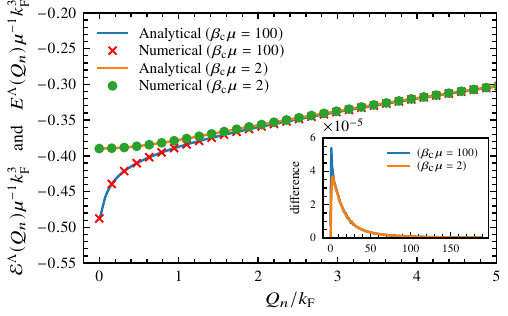}
  \caption{(Color online) The markers indicate the  numerically calculated cut-off dependent eigenvalues, $E^{\Lambda}(Q)$ at  discrete $Q_n$,  of the half-slab pairing
  kernel $\curK^{\Lambda}$,  Eq.~\eqref{eq:eigenvaluerelation}, for representative high and low  $T_{\rm c}$ values (note $\beta_{\rm c} = (k_{\rm B}T_{\rm c})^{-1})$. The solid lines show the cut-off dependent bulk eigenvalues ${\cal E}^{\Lambda}(Q)$, Eq.~\eqref{eq:epsilonLambdaQ}, at the same temperatures. The inset shows the difference $|{\cal E}^{\Lambda}(Q)-E^{\Lambda}(Q)|$ between the half-slab eigenvalues and the bulk values. To numerical precision they agree. The critical coupling at $T_{\rm c}$ is $g_{\rm c}= 1/{\cal E}^{\Lambda}(0)$.}
  \label{fig:Eigenvalues}
\end{figure}
The corresponding eigenvalues $E^\Lambda(q)$  of the kernel $\curK^\Lambda$ are found to be 
equal, within precision,
to $\curE^\Lambda(q)$, as shown in 
Fig.~\ref{fig:Eigenvalues}, consistent with Weyl's
theorem discussed above. Thus, our numerics confirms that  the final term  $-\frac{1}{2} f^{\Lambda}_{Q,Q'}$ in Eq.~(\ref{Eq:kappamatrix}) does not modify
the essential spectrum of the kernel, which is controlled
by the singular first term.  
We note that Weyl's theorem does not exclude 
the possibility of discrete bound states, 
which (if present) could imply the presence of
a surface pairing instability preceding the bulk.
However, numerically we find no evidence of any bound states of $\curK^\Lambda$, suggesting that the spectrum of this 
kernel is given only by the continuum spectrum 
$\curE^\Lambda(q)$ with $q>0$.
\begin{figure}[t]
  \centering
  \includegraphics[width=\columnwidth]{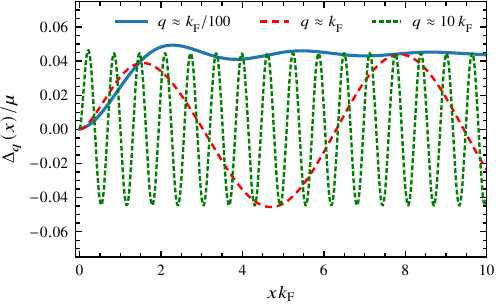}
  \caption{(Color online) Shown here are representative  eigenstates of the $T_{\rm c}$-equation, Eq.~\eqref{eq:eigenvaluerelation}, in real space for three
  values of the wavevector quantum number $q$ and at low $T_{\rm c}$   ($\beta\mu = \beta_\mathrm{c}\mu=100$).  While the higher-$q$ states look basically like $\sin(qx)$ (as seen in the green short-dashed and red dashed curves), the low-$q$ state (solid blue) exhibits additional Friedel oscillations for $x\to 0$.  }
  \label{fig:LowTcEigenstates}
\end{figure}

Although the spectrum is independent of the second
term of $\curK^\Lambda$, the eigenfunctions
do depend on this term.  This can be seen
in their real-space form, which is
obtained by the cosine transform formula,
\be
\Delta_{q}(x) = \int_0^\infty \frac{{\rm d}Q}{\pi} \chi_Q(x) 
\Delta_{q}(Q).
\label{eq:realspaceef}
\ee
In Fig.~\ref{fig:LowTcEigenstates}, we show
typical results for the real-space eigenfunctions of $\curK^\Lambda$ at low $T$ (corresponding to the 
weak-coupling BCS regime), obtained using 
Eq.~(\ref{eq:realspaceef}).  We show three different
eigenfunctions with wavevector quantum numbers
$q = \kf/100$, $q= \kf$ and $q \approx 10\kf$. We
see that the higher-$q$  (red dashed and green short-dashed) curves look, essentially, like
$\sin(q x)$.  The lowest $q$ curve (blue), however,
has additional oscillations at the scale $2\kf$, 
although a \lq\lq zoomed-out\rq\rq\ plot of this curve would show that it also oscillates at the long
wavevector $\kf/100$.

The  $2\kf$ Friedel oscillations in the 
BCS regime eigenfunctions are  suppressed with  increasing coupling (or equivalently, increasing transition temperature) as we move
towards the stronger coupling regime.
\begin{figure}[t]
  \centering
  \includegraphics[width=\columnwidth]{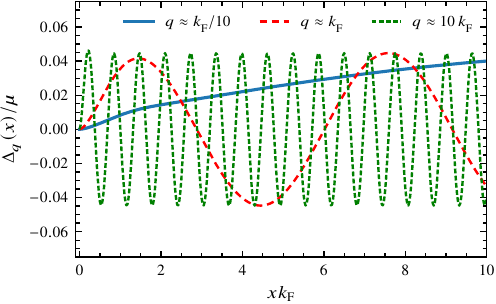}
  \caption{(Color online) Shown here are representative  eigenfunctions of the $T_{\rm c}$-equation, Eq.~\eqref{eq:eigenvaluerelation}, in real space for 
  three values of the wavevector quantum number $q$ in the near-unitary (or BEC) regime with 
  $\beta\mu = \beta_\mathrm{c}\mu=2$.
  Like in Fig.~\ref{fig:LowTcEigenstates}, the
  higher $q$ states (green short-dashed and red dashed curves) look basically like $ \sin(qx)$.
  However, the low $q$ case (solid blue curve) does not exhibit the additional Friedel oscillations seen in Fig.~\ref{fig:LowTcEigenstates}. Figure \ref{fig:HighTcEigenstates for very small q0} further highlights the coupling dependence of eigenstates at  very small $q$. }
  \label{fig:HighTcEigenstates}
\end{figure}
\begin{figure}[t]
  \centering
  \includegraphics[width=\columnwidth]{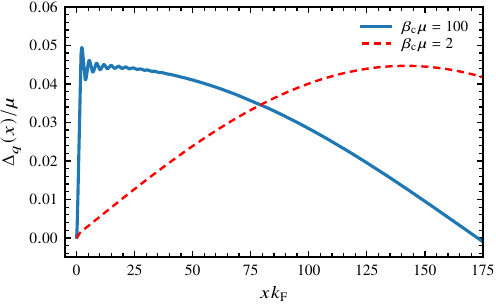}
  \caption{(Color online) Shown here is a  comparison of the
  kernel eigenstates at low $q$ (i.e.,  $q \approx k_{\rm F}/100$) in the weak-coupling BCS regime (blue solid curve) and strong coupling near-unitarity regime (red dashed curve).  The BCS curve exhibits strong edge effects at the scale 
  $\kf^{-1}$ (which we'll find persist below $\tc$), while
  the BEC curve shows almost no edge effects. Instead,  it  simply looks like
  $\sin(qx)$.}
  \label{fig:HighTcEigenstates for very small q0}
\end{figure}
This is illustrated in Fig.~\ref{fig:HighTcEigenstates} which shows
three typical eigenfunctions with different
$q$ values for the case of a stronger coupling
value towards the BEC regime.  In this figure, 
all three curves look essentially like 
$\sin(qx)$, without any edge Friedel oscillations, although the lowest $q$ case shows a slight kink near $x\to 0$.  

To further illustrate the difference,
in Fig.~\ref{fig:HighTcEigenstates for very small q0} we compare the 
eigenfunctions at the same small 
$q\approx \kf/100$ for two different 
coupling values.   Although the 
long distance oscillations are at the same wavelength,
the low-$T_{\rm c}$ curve (blue solid) exhibits
Friedel oscillations and a rapid drop,
while the high-$T_{\rm c}$ curve (red dashed) 
does not.

\subsection{Analytical analysis}
To obtain an analytical understanding of
the behavior of the eigenfunctions, 
and directly show that the continuum eigenvalues are 
equal to the bulk eigenvalues,
we 
consider a simple ansatz that solves
the relevant eigenvalue equation (with eigenvalue $E^\Lambda(q)$):
\be
\label{eq:eigenvaluerelation}
\int_0^\infty \frac{{\rm d}Q'}{\pi} \langle Q|\curK^\Lambda|Q'\rangle
\Delta_{q}(Q') = E^\Lambda(q) \Delta_{q}(Q).
\ee
The ansatz for solving this equation consists
of a delta-function piece plus a correction:
\be
\label{Eq:deltavsQN}
\Delta_{q}(Q) = \pi\delta(Q-q) -\Delta_{q}^{(1)}(Q),
\ee
where we recall that  capital $Q$ refers to the wavevector
argument of the  function, while $q$ is the
wavevector quantum
number.  

If we assume the corresponding real space eigenfunctions 
satisfy $\Delta_{q}(x)\big|_{x\to 0} =0$, then
plugging into Eq.~(\ref{eq:realspaceef})
leads to the condition
\be
\label{Eq:delta1Norm}
\int_0^\infty  \frac{{\rm d}Q}{\pi}  \Delta_{q}^{(1)}(Q) =1,
\ee
on the function $\Delta_{q}^{(1)}(Q)$.
Next, we plug the ansatz Eq.~(\ref{Eq:deltavsQN}) into 
the eigenvalue equation Eq.~(\ref{eq:eigenvaluerelation})
and simplify to get 
\bea
&& \pi \curE^\Lambda(q) \delta(Q-q) - \curE^\Lambda(Q) \Delta_{q}^{(1)}(Q) 
- \frac{1}{2}f_{Q,q}^\Lambda 
\\
\nonumber &&
 \qquad \qquad + \frac{1}{2}\int_0^\infty \frac{{\rm d}Q'}{\pi} f_{Q,Q'}^\Lambda
\Delta_{q}^{(1)}(Q') 
 \\
\nonumber &&\qquad
 = E^\Lambda(q) \Big[\pi \delta(Q-q) - \Delta_{q}^{(1)}(Q)\Big].
\eea
Note the presence of the delta-function distributions on the left and right. 
Since no other such distributions are present, for this equation to be satisfied the eigenvalue must satisfy 
$E^\Lambda(q)= \curE^\Lambda(q)$.
This confirms the result we found numerically, and consistent with 
Weyl's theorem, that
the continuous kernel eigenvalues are equal to the eigenvalues of the singular
\lq\lq diagonal\rq\rq\ part of the kernel.  

Using this, the integral equation
for $\Delta_q^{(1)}(Q)$ simplifies to:
\bea
\nonumber
&&\hspace{-0.5cm}\Big(  \curE^\Lambda(q)-\curE^\Lambda(Q)\Big) \Delta^{(1)}_{q}(Q) 
\\
&&\qquad \qquad =
\frac{1}{2}\Big(f_{Q,q}^\Lambda -
\int_0^\infty \frac{{\rm d}Q'}{\pi} f_{Q,Q'}^\Lambda 
\Delta^{(1)}_{q}(Q')\Big)
\nonumber
\\
 &&\qquad \qquad =
\frac{1}{2}
\int_0^\infty \frac{{\rm d}Q'}{\pi} \Big(  f_{Q,q}^\Lambda-f_{Q,Q'}^\Lambda\Big)
\Delta^{(1)}_{q}(Q'),
\label{Eq:bothsatisfy2}
\eea
where to get to the last line we used 
the condition Eq.~(\ref{Eq:delta1Norm}).
  Recall that the functions $\curE^\Lambda(Q)$ and 
$f_{Q,Q'}^\Lambda$ are both cutoff dependent since the
integrals defining them are divergent for $\Lambda\to 
\infty$.  However,
in Eq.~(\ref{Eq:bothsatisfy2}) these functions
both enter as differences, i.e., 
$\curE^\Lambda(q)-\curE^\Lambda(Q)$ and
$f_{Q,q}^\Lambda-f_{Q,Q'}^\Lambda$ that 
are cutoff independent.  Thus,
 as in Eq.~(\ref{eq:combinedintegral}), 
we can combine the integrals defining them  and take the limit of $\Lambda \to \infty$. 
For example, we explicitly have 
\bea
&&f_{Q,q}^\Lambda-f_{Q,Q'}^\Lambda
=\int
\frac{{\rm d}^2 k_\parallel}{(2\pi)^2}
\Big[ \frac{\nf(\xi_{k_{\parallel} k_1}) - \nf(-\xi_{k_{\parallel} k_2})}
     {\xi_{k_{\parallel} k_1}+\xi_{k_{\parallel} k_2}}
\nonumber
     \\
  && 
\qquad \qquad 
- \frac{\nf(\xi_{k_{\parallel} k_3}) - \nf(-\xi_{k_{\parallel} k_4})}
     {\xi_{k_{\parallel} k_3}+\xi_{k_{\parallel} k_4}}\Big],     
\eea
where $k_1 = \frac{1}{2}(Q-q)$, $k_2 = \frac{1}{2}(Q+q)$, $k_3 = \frac{1}{2}(Q-Q')$, and
$k_4 = \frac{1}{2} (Q+Q')$, a convergent integral.
Similar steps show that $\curE^\Lambda(q)-\curE^\Lambda(Q)$ can be written in a cutoff-independent way.  
This tells us that while the eigenvalues of the pairing kernel are cutoff dependent, the eigenfunctions are cutoff independent in the 
limit of $\Lambda\to \infty$, i.e., they are universal.

In fact, the cutoff dependence of the eigenvalues can
be simply renormalized using the same procedure
as in the Lippmann-Schwinger equation, Eq.~(\ref{lambdaasrelation}).
We define the renormalized energy:
\bea
\label{Eq:RenormEnergy}
\curE^R(q) &\equiv &  \curE^\Lambda(q) + \int_\Lambda \frac{{\rm d}^3 k}{(2\pi)^3} \frac{1}{2\epsilon_k} ,
\\
&=&\int \frac{{\rm d}^3 k}{(2\pi)^3} \Big(
\frac{\nf(\xi_{\bk+\frac{1}{2}\bq}) - \nf(-\xi_{\bk+\frac{1}{2}\bq})}
{\xi_{\bk+\frac{1}{2}\bq}+\xi_{\bk-\frac{1}{2}\bq}} + \frac{1}{2\epsilon_k}\Big),
\nonumber 
\eea
where in the second line we combined the cutoff-dependent
integrals and subsequently set the cutoff to infinity.
In this step we also combined the $k$ and $Q$ integrations into a single three-dimensional $k$ 
integral.
Here, 
$\bq$ is a wavevector along the $\xh$ direction of length
$q$.  In
the limit of $q \to 0$, $\curE^R(q)$ has the form
\bea
\curE^R(q) &\simeq & \frac{1}{g(T)} + \frac{1}{2}\rho q^2,
\\
\frac{1}{g(T)} &\equiv & - \int \frac{{\rm d}^3 k}{(2\pi)^3} 
\Big( \frac{\tanh\frac{\xi_k}{2T}}{2\xi_k} - \frac{1}{2\epsilon_k}\Big),
\nonumber 
\eea
where $\frac{1}{g(T)}$ determines the bulk $\tc$.  That is, for a given renormalized coupling $g$ the bulk $\tc$ is given 
by the solution to $\frac{1}{g} = \frac{1}{g(T_c)}$.  The coefficient
$\rho$ can be obtained by Taylor expanding the integrand 
of the second line of Eq.~(\ref{Eq:RenormEnergy}) in small $q$.
In the weak-coupling (low $T$) limit an approximate
analytic form for $\rho$ follows from converting the $k$ integral to an
energy integral, approximating the density of states by its value at the 
Fermi energy, and extending the integration to $-\infty$ to obtain 
\be
\rho \simeq \frac{m\kf}{2\pi^2 }
\frac{7\zeta(3)\vf^2}{24\pi^2 T^2},
\ee
which
precisely agrees with the well-known Ginzburg-Landau result.

Now we return to Eq.~(\ref{Eq:bothsatisfy2}).  The preceding steps explicitly show
that the quantities $\curE^\Lambda(q) - \curE^\Lambda(Q)$ and $f_{Q,q}^\Lambda-f_{Q,Q'}^\Lambda$ are cutoff-independent.
Nonetheless for notational simplicity we'll keep the cutoff superscript.  To 
simplify Eq.~(\ref{Eq:bothsatisfy2}),
we
 define $\delta_{q}(Q)$ via,
 \be
 \label{Eq:bigdlittled}
\Delta_{q}^{(1)}(Q) = \curP \frac{\delta_{q}(Q)}{\curE^\Lambda(q)-\curE^\Lambda(Q)},
 \ee
 where $\curP$ indicates that, when integrating, we'll 
 interpret the singularity at $Q\to q$ using a principal-value prescription, 
 i.e., via $\curP\frac{1}{Q-q} = {\lim}_{\epsilon\to 0}
 \frac{Q-q}{(Q-q)^2+\epsilon^2}$.  
 Clearly, the singular behavior shown in
 Fig.~\ref{fig:LowTcEigenstates in Momentum space} for $Q\to q$ is precisely reflected 
 in Eq.~(\ref{Eq:bigdlittled}).  
 
 With the definition Eq.~(\ref{Eq:bigdlittled}), 
  Eq.~(\ref{Eq:bothsatisfy2}) reduces to
 an integral equation for $\delta_{q}(Q)$:
 \be
 \delta_{q}(Q) = \frac{1}{2}\int_0^\infty \frac{{\rm d}Q'}{\pi}
 \frac{f_{Q,q}^\Lambda-f_{Q,Q'}^\Lambda}{\curE^\Lambda(q)-\curE^\Lambda(Q')}
 \delta_{q}(Q'),
 \label{Eq:eomITOlittled}
 \ee
where now the integrand is nonsingular for $Q'\to q$ and also independent of the cutoff  $\Lambda$ as we have discussed.
Numerically, it is straightforward
to solve for $\delta_{q}(Q)$, and 
the numerical results we find precisely
agree with the results of direct
numerical 
diagonalization of the full kernel.

Combining the preceding results, we have the following final form for the eigenfunctions of the kernel 
\be
\Delta_{q}(x) = \int_0^\infty \frac{{\rm d}Q}{\pi} \frac{\delta_{q}(Q)}
      {\curE^\Lambda(q)-\curE^\Lambda(Q)}
      \Big(\chi_{q}(x) - \chi_Q(x)\Big).
      \label{eq:unnormeigen}
      \ee
      To use these as a basis set in the half-space, we must correctly normalize these
      functions.  

\section{Eigenfunction Normalization}
\label{sec:eeignorm}
In the preceding section, we derived Eq.~(\ref{eq:unnormeigen}) for the pairing kernel eigenfunctions, where $\delta_{q}(Q)$ is determined by Eq.~(\ref{Eq:eomITOlittled}). 
To normalize these eigenfunctions, we need the inner product 
\be
\int_0^\infty {\rm d}x \, \Delta_{q}(x) \Delta_{q_1}(x)  = 
\int_0^\infty \frac{{\rm d}Q}{\pi} \, \Delta_{q}(Q) \Delta_{q_1}(Q).
\ee
Following standard arguments, this
 inner product must vanish for $q\neq q_1$, 
 since the factors correspond to eigenfunctions of $\curK^\Lambda$ with different eigenvalues. 
 Therefore, we conclude that this inner product is 
 proportional to a delta function $\delta(q -q_1)$.
To get the coefficient, we evaluate the integral using Eq.~(\ref{Eq:deltavsQN}), getting
\bea
\label{Eq:twolines}
&&\int_0^\infty 
\frac{{\rm d}Q}{\pi}\, \Delta_{q}(Q) \Delta_{q_1}(Q) = \pi\delta(q - q_1)
\\
&&
 + \int_0^\infty \frac{{\rm d}Q}{\pi}
 \Delta_{q}^{(1)}(Q)\Delta_{q_1}^{(1)}(Q)
 -\Delta_{q}^{(1)}(q_1)-\Delta_{q_1}^{(1)}(q).
 \nonumber 
\eea
From the above arguments, we know the second line must vanish for
$q\neq q_1$.  As we now show, there is a 
subtle additional delta function
contribution coming from the integral in
the second line.  To see this, we note that
the denominator of
Eq.~(\ref{Eq:bigdlittled}) vanishes for $Q\to q$,
so that $\Delta_{q}^{(1)}(Q)$ has the singular behavior
 \be
 \label{eq:deltaAsymp}
\Delta_{q}^{(1)}(Q)  \simeq \curP\frac{\curR(q)}{Q-q} , \text{for} 
\,\,\,Q\to q,
 \ee
 with $\curR(q)$ the residue, given by
 \be
\curR(q) = -\frac{\delta_{q}(q)}{\curE^{\Lambda '}(q)},
\label{Eq:curRresult}
 \ee
 assuming that the numerator does not vanish (which we 
 always find to be the case numerically). Here the prime
 denotes differentiation with respect to the argument.  
Note that the integral in the second line of 
Eq.~(\ref{Eq:twolines}) contains a product of two such
principal value factors at $q$ and $q_1$, which give
an additional delta-function contribution for $q \to q_1$, i.e., 
a term proportional to $\delta(q-q_1)$.
In Appendix~\ref{app:dvpi}, we review the fact that the integral of the product of
two principal value factors yields a delta function. To get the 
prefactor  of the delta-function contribution, it is clear (since we have already
established that the second line of Eq.~(\ref{Eq:twolines}) vanishes for $q \neq q_1$)
that all that matters is the behavior of the integrand in
the vicinity of $q$ and $q_1$, i.e., the residues $\curR$. 
Including all contributions then finally gives:
\be
\int_0^\infty 
\frac{{\rm d}Q}{\pi}\, \Delta_{q}(Q) \Delta_{q_1}(Q)
=\pi \Big[1+\big(\curR(q)\big)^2
    \Big]\delta(q-q_1),
    \label{Eq:deltainnerQ}
\ee
where the second term in square
brackets came from the 
second line of Eq.~(\ref{Eq:twolines}).
This tells us that the proper delta-function
normalized cosine-basis eigenfunctions are 
\be
\Deltab_{q}(Q) = \frac{1}{\sqrt{1+\big(\curR(q)\big)^2}}
\Delta_{q}(Q).
\label{Eq:deltabNorm}
\ee
The corresponding  real-space expression is:
\be
\Deltab_{q}(x) = 
\frac{1}{\sqrt{1+\big(\curR(q)\big)^2}}
\Big( \chi_{q}(x)  - \int_0^\infty \frac{{\rm d}Q}{\pi}
\chi_Q(x)\Delta^{(1)}_{q}(Q)\Big).
\label{eq:deltab}
\ee
We can use this expression to study the large-$x$
behavior of the eigenfunctions. At large $x$, we need
the behavior of $\Delta^{(1)}_{q}(Q)$ for small
$Q$, where we know it has a pole at $Q= q$.  We therefore
separate out this piece by writing 
\be
  \label{Eq:delta1delta2}
  \Delta_{q}^{(1)}(Q) = \curP\frac{2q \curR(q)}{Q^2-q^2} +
  \Delta_{q}^{(2)}(Q),
  \ee
  where, by definition, $\Delta_{q}^{(2)}(Q)$ has no pole at $q$.  
  Upon plugging this
  into Eq.~(\ref{eq:deltab}), we can evaluate the 
  contribution coming from the first term of 
  Eq.~(\ref{Eq:delta1delta2}) to get:
  \bea
  \nonumber 
&&  \Deltab_{q}(x) =\frac{1}{\sqrt{1+\big(\curR(q)\big)^2}}
\Big(\chi_{q}(x) + \curR(q) \psi_{q}(x)
\\
&& \qquad \qquad 
  - \int_0^\infty \frac{{\rm d}Q}{\pi} \chi_Q(x)\Delta_{q}^{(2)}(Q) 
  \Big),
  \label{Eq:deltab}
  \eea
  an alternate expression for $\Deltab_{q}(x)$
  showing the large-$x$ behavior (or asymptotic scattering form) in which the
  eigenfunctions are a sum of sine and cosine
pieces (with relative weight determined by $\curR(q)$), plus
the final third term, coming from $\Delta_{q}^{(2)}(Q)$, that modifies the eigenfunction shape 
near $x\to 0$.  

We note here some additional interesting properties of 
these eigenfunctions.  
Firstly, since the eigenvalues 
$\curE^\Lambda(q)$
have a quadratic dependence on $q$ for small $q$,
Eq.~(\ref{Eq:curRresult}) implies that $\curR(q)
\propto\frac{1}{q}$ for small $q$ (assuming 
$\delta_{q}(q)$ is finite in the limit of
$q\to 0$, which we find to be true numerically).
This furthermore implies that, at small $q$, 
the normalized eigenfunctions are approximately
given by $\Deltab_{q}(x) \approx \psi_{q}(x)$, i.e.,
they are sine eigenfunctions, which formally
vanish for $q = 0$ (i.e., the eigenfunctions 
are defined for $q>0$ only, an aspect we discuss further below).  

Secondly, if  one naively Taylor
expands the argument of the final integral, 
Eq.~(\ref{Eq:deltab}) seems to imply that 
 $\Deltab_{q}(x)$ vanishes linearly in $x$ for 
small $x$. In fact, such a Taylor expansion for small
$x$ is not valid in this integral, due to the fact that 
$\Delta_{q}^{(2)}(Q)\propto 1/Q^2$ for large $Q$.  
A Taylor expansion is, however, valid for 
Eq.~(\ref{eq:deltab}), implying the eigenfunctions 
satisfy $\Deltab_{q}(x)\propto x^2$ for small $x$ 
(which we also find numerically).

Although orthogonality of the eigenfunctions, 
\be
\int_0^\infty {\rm d}x \, \Deltab_{q}(x) \Deltab_{q_1}(x)
=\pi\delta(q-q_1),
\ee
follows from the preceding analysis,
we have not directly proven completeness.
Indeed, the generalized eigenfunctions we have found are the analog of scattering states in
quantum mechanics, with a continuous spectrum.
Following results in spectral theory, in principle there may be bound states of the kernel, i.e., 
square normalizable solutions with eigenvalues
below the continuum.  These would correspond to
"surface" pairing instabilities of the half-slab
system.

Although such bound states are possible,
and would be needed in the correct generalized completeness relation for the kernel eigenfunctions, we find no evidence for them numerically.  Indeed,
  numerical analysis of the kernel (via approximating it by a square matrix) finds
  only
solutions that are consistent with the
continuum eigenfunctions Eq.~(\ref{eq:deltab}) discussed above. Excluding such
bound states analytically is a
difficult task. To do this, one strategy is to derive a bound on the spectrum of $\curK^\Lambda$.
If, for example, one could show this spectrum is bounded from below by $\curE^\Lambda(0)$, this would preclude any bound states below the continuum spectrum. This would preclude any bound states at all, if we can apply the common assumption that one 
does not expect any bound states in the continuum. Although we have not succeeded to analytically
show such a bound, in Appendix~\ref{app:bound} we demonstrate the weaker bound 
$E^{\Lambda}\geq 2\curE^\Lambda(0)$ on any eigenvalues 
$E^{\Lambda}$ of $\curK^\Lambda$.  

Applying our numerical finding of no bound states of $\curK^\Lambda$, 
the absence of bound states implies that the continuum eigenfunctions form
a complete generalized basis.  We 
 then expect the continuum eigenstates $\Deltab_q(x)$  to obey   
 a distributional completeness relation of the form:
\be
 \int_0^\infty \frac{{\rm d}q}{\pi}
  \Deltab_{q}(x) \Deltab_{q}(x') =  \delta(x-x').
  \label{eq:completenessDeltaQ}
\ee
This  also implies we can express the cutoff-dependent kernel in terms of
the eigenfunctions:
  \be
  \label{eq:kerneleigenfunctionexpansion}
  \curK^\Lambda(x,x')  = \int_0^\infty \frac{{\rm d}q}{\pi}
  \curE^\Lambda(q) \Deltab_{q}(x) \Deltab_{q}(x') .
  \ee
As we have argued, only the eigenvalues are cutoff-dependent, with 
the eigenfunction shape being universal for $\Lambda\to \infty$.  
And, since the energy renormalization, Eq.~(\ref{Eq:RenormEnergy}),
is identical to the coupling renormalization Eq.~(\ref{lambdaasrelation}), 
the linearized cutoff-dependent $\tc$ equation, 
$\frac{1}{\lambda}\Delta(x) = \int_0^\infty {\rm d}x'\, \curK^\Lambda(x,x')\Delta(x')$,
can be rewritten in the renormalized form (using 
Eq.~(\ref{eq:kerneleigenfunctionexpansion})):
\bea
\label{Eq:linearizedRenormalized}
\frac{1}{g}\Delta(x) &=& \int_0^\infty {\rm d}x' \curK^R(x,x')\Delta(x'),
\\
\curK^R(x,x') &\equiv & \int_0^\infty \frac{{\rm d}q}{\pi} 
\curE^R(q) \Deltab_{q}(x) \Deltab_{q}(x') ,
\eea
giving the renormalized linear $\tc$ equation and the renormalized 
pairing kernel $\curK^R(x,x')$.  

\section{Edge pairing for $T<\tc$}
\label{sec:edgepairing}
Our numerical and analytical results from the above tells us
that the generalized eigenfunctions of the pairing kernel $\curK^\Lambda$
have the lowest eigenvalue, $\curE^\Lambda(0)$, that is equal
to the bulk value.  This tells us that, within our approach, the transition temperature of the half slab is equal to that of
the bulk system.  However, for low $q$ the eigenfunctions
behave as $\propto \sin(q x)$, vanishing as $q \to 0$.  This means that we cannot apply the conventional wisdom of assuming that the lowest eigenfunction of the pairing kernel determines the shape of the pairing slightly below $\tc$.  

In fact, the same phenomenon occurs in standard Ginzburg-Landau (GL)
theory: for a semi-infinite system with 
Dirichlet boundary condition at $x=0$, the eigenfunctions 
of the GL kernel (which, up to a constant, is proportional to the Laplacian operator) are exactly
$\psi_k(x) = \sqrt{2}\sin(kx)$.  
However, the actual shape of the 
pairing amplitude 
below $\tc$ in GL theory (of the form of
a hyperbolic tangent function) comes from solving the 
full nonlinear problem below $\tc$. In Appendix~\ref{app:GL}, we review these aspects of GL theory and describe a simple approximate method to understand edge effects in the full nonlinear GL theory.

The lessons from GL theory for our present problem 
are twofold.  Firstly, as noted above, while the conventional wisdom
is  that one can simply take the eigenfunction of the pairing kernel with the lowest eigenvalue as a proxy for the pairing near $\tc$, that 
procedure does not work for the half-slab problem.
Secondly, the spatial structure of the pairing kernel eigenfunctions may not tell us anything about the full
nonlinear problem.  However, as we show below, in the present case we find that the Friedel-oscillation structure of the pairing eigenfunctions at weak coupling are indeed
reflected in the nonlinear problem below $\tc$.  

To demonstrate this, we now analyze pairing below $\tc$ in a semi-infinite system.  To do this, we add a GL-inspired nonlinear term to the pairing
problem Eq.~(\ref{Tc eigenvalue equation in real space}), i.e., we study:
\be
\frac{1}{\lambda}\Delta(x)= \curK^\Lambda\Delta(x) + \beta \Delta(x)^3.
\label{eq:fullnonlinear}
\ee
Here, $\curK^\Lambda\Delta(x) \equiv
\int_0^\infty {\rm d}x' \, \curK^\Lambda(x,x')\Delta(x')$ is a shorthand
for the full kernel acting on $\Delta(x)$ and the parameter $\beta$ controls the 
magnitude of $\Delta(x)$ at large $x$.
We first present an approximate ``linearized''
method to study Eq.~(\ref{eq:fullnonlinear}),
before presenting our full numerical
solution below.

\begin{figure}[t]
  \centering
  \includegraphics[width=\columnwidth]{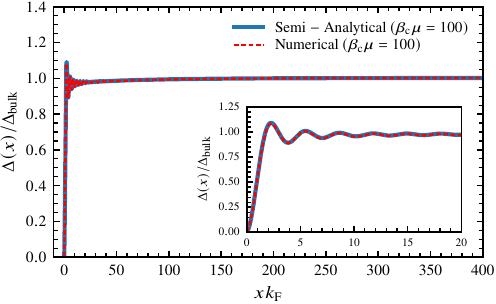}
  \caption{(Color online) The dashed red line shows the spatial profile of the local pairing $\Delta(x)$, normalized to the bulk value, for the full numerical solution of the nonlinear problem in the regime $T<T_{\rm c}$.  Here, we chose a low
  $\tc$ (i.e., $\beta_{\rm c}\mu = 100$ or $ \kf \as \approx - 0.38$) and temperature $\beta\mu = 1.1 \beta_\mathrm{c}\mu$ or $(T_{\rm c}-T)/T_{\rm c}\approx 0.09$. The  solid blue line shows the approximate semi-analytical results at the same temperature. The inset shows local pairing closer to the edge. The Friedel-like oscillations occur with a wavelength  of approximately $2k^{-1}_{\rm F}.$ At this low $T_{\rm c}$ the full numerical and semi-analytical approach agree very well. Starting from $x=0$, the local pairing rises very quickly (accompanied by Friedel oscillations) and  approaches the bulk value on a length scale of approximately $2k^{-1}_{\rm F}$. At larger $x$, $\Delta(x)$ slowly relaxes to its bulk value on a length scale determined by the coherence length  $\xi(T)\gg k^{-1}_{\rm F}$.   }
  \label{fig:LowTc Delta}
\end{figure}
\begin{figure}[t]
  \centering
  \includegraphics[width=\columnwidth]{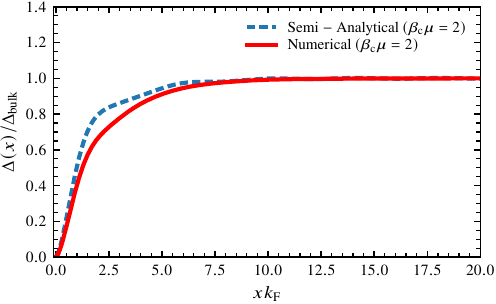}
  \caption{(Color online)  The solid line shows the spatial profile of the local pairing $\Delta(x)$, normalized to the bulk value, for the full numerical solution of the nonlinear problem at $T<T_{\rm c}$.  Here we chose a relatively high $\tc$ 
   $(\beta_{\rm c}\mu = 2$ or $\kf \as \approx - 6.1)$ and temperature $\beta\mu = 1.1 \beta_\mathrm{c}\mu$ or $(T_{\rm c}-T)/T_{\rm c}\approx 0.09$. This $T_{\rm c}$ corresponds to a coupling that is approaching the unitary regime.  In comparison the  dashed line shows the approximate semi-analytical results.  At this high $T_{\rm c}$ the differences between the full numerical and semi-analytical approximation become much more apparent. In addition,  
   Friedel-like oscillations in the local pairing have been almost completely washed away, with the pairing
   showing a slow relaxation to the bulk value (qualitatively
   consistent with Ginzburg-Landau theory).
   }
  \label{fig:HighTc Delta}
\end{figure}

\subsection{Semi-analytical linearized solution}
\label{sec:sals}
Our main goal is to study the behavior of
the pairing amplitude near a hard wall
in the regime $T\alt \tc$. Our linearized approximation is inspired by a similar
approximation one can make in conventional
GL theory (see Appendix~\ref{app:GL}). We write
\be
\Delta(x) = \Delta_0 +\Delta_1(x),
\label{eq:expansionDelta1}
\ee
with
$\Delta_0$ the bulk value (which we also call
$\Delta_{\rm bulk}$ below and in figures) at $x\to \infty$, which is determined by
the renormalized
energy at zero wavevector, i.e.,
\be
\frac{1}{g}\Delta_0 = \curE^R(0) \Delta_0 +\beta\Delta_0^3 .  
\ee
Using $\curE^R(0) = \frac{1}{g(T)}$, we get
\be
\label{eq:needbelowDeltanought}
\Delta_0 = \sqrt{\frac{1}{\beta}\left( \frac{1}{g} - \frac{1}{g(T)}\right)}.
\ee
To get an approximate equation for the correction, $\Delta_1(x)$, we plug 
Eq.~(\ref{eq:expansionDelta1}) into 
Eq.~(\ref{eq:fullnonlinear}), Taylor expand all terms to
leading (linear) order in small $\Delta_1(x)$, then re-express 
in terms of $\Delta(x)$.  The result is:
\be
\label{eq:approxNonlinearEquation}
\curK^R\Delta(x)  = \left( \frac{1}{g} - 3\beta \Delta_0^2
\right) \Delta(x) + 2\beta\Delta_0^3 .
\ee
Now that we have a linear equation, to proceed we
 expand all terms in the basis set of the kernel eigenfunctions
$\Deltab_{q}(x)$.  Thus, we assume the solution is of
the form 
\be
\Delta(x) = \int_0^\infty \frac{{\rm d}q}{\pi}\Deltab_q(x) \Delta_q,
\label{eq:deltabartransform}
\ee
with $\Delta_q$ being unknown coefficients that we aim to find.  Essentially, we are transforming our 
equation with respect to the basis set $\Deltab_q(x)$.
Since the last term on the right side of Eq.~(\ref{eq:approxNonlinearEquation})
is constant (i.e., $x$ independent), to accomplish this
transform we need the coefficients $\Delta_{q}^{\bone}$
that satisfy
\be
1 = \int_0^\infty \frac{{\rm d}q}{\pi} \Deltab_{q}(x) \Delta_{q}^{\bone}.
\ee
Direct calculation shows that $\Delta_{q}^{\bone} =
\frac{1}{\sqrt{2}}\lim_{Q\to 0} \Deltab_{q}(Q) 
$, i.e., the coefficients are given by the cosine-basis eigenfunctions 
at wavevector $Q\to 0$.  Then, the transform of Eq.~(\ref{eq:approxNonlinearEquation})
with respect to the basis set $\Deltab_{q}(x)$ is:
\be
\curE^R(q) \Delta_{q} = \Big( \frac{1}{g} - 3\beta \Delta_0^2
\Big) \Delta_{q} + 2\beta\Delta_0^3 \Delta_{q}^{\bone}.
\ee
Solving for the coefficients $\Delta_{q}$, we get
\be
\Delta_{q}
= \frac{2\beta\Delta_0^3 \Delta_{q}^{\bone}}{\curE^R(q) -\curE^R(0)  + 2\beta \Delta_0^2},
\label{eq:deltaqsemianalytical}
\ee
where to simplify we used Eq.~(\ref{eq:needbelowDeltanought}) as well as $\frac{1}{g(T)} = \curE^R(0)$.

In Figs.~\ref{fig:LowTc Delta} 
and \ref{fig:HighTc Delta} we compare
the full numerical solution to
Eq.~(\ref{eq:fullnonlinear}) (discussed in the subsequent subsection) to this semi-analytical 
approach, with the latter curves obtained by 
plugging Eq.~(\ref{eq:deltaqsemianalytical}) 
into Eq.~(\ref{eq:deltabartransform})
to get  $\Delta(x)$.  
We see
that, although approximate, it works relatively well in both the weak coupling BCS regime (Fig.~\ref{fig:LowTc Delta}) and in the strong
coupling
near unitarity regime (Fig.~\ref{fig:HighTc Delta}).  

\subsection{Full numerical solution}

\begin{figure}[t]
  \centering
  \includegraphics[width=\columnwidth]{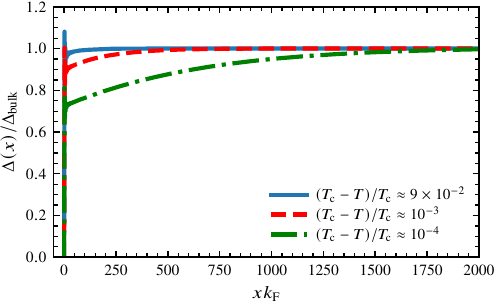}
  \caption{(Color online) The three curves illustrate the temperature evolution of 
  $\Delta(x)$  at low $T_{\rm c}$ ($\beta_{\rm c}\mu =100$) as $T\to T_{\rm c}$, all normalized to their bulk values.  The blue solid curve shows the lowest $T$, while the red dashed and green dot-dashed show the behavior for $T$ progressively closer to $\tc$.  
  Here we have zoomed out to 
  emphasize the long-distance behavior, showing a spatial variation on long length scales consistent
  with Ginzburg-Landau theory. Near $x\to 0$ these
  curves exhibit Friedel oscillations and a 
  sharp drop to zero like in Fig.~\ref{fig:LowTc Delta}.}
\label{fig:TemperatureEvolutionAtLowTc}
\end{figure}

To fully understand the pairing below
$\tc$ within our model, we must
numerically analyze Eq.~(\ref{eq:fullnonlinear}).  Our aim is to
understand how (and whether) the edge behavior of
the eigenstates of the pairing kernel are
reflected in the edge pairing as 
temperature is reduced below the transition.
We reserve many technical details of our method to Appendix~\ref{app:numdet},
only emphasizing that our method does
utilize the kernel eigenstates as a 
basis set.  

We start in the low-$\tc$ weak coupling BCS regime, with typical results shown in Fig.~\ref{fig:LowTc Delta}, for the case of 
$T/\tc\simeq 0.91$.  At this 
temperature, the system is far 
enough below $\tc$ that the bulk
pairing is basically spatially 
uniform, with only a sharp drop
near $x=0$.  
(Note
these results, and subsequent results,
are always plotted normalized to the
bulk value $\Delta_{\rm bulk}$ of the pairing, which vanishes
for $T\to \tc$.)
The pairing 
shown in this figure is almost exactly of
the form $\Delta(x) = \Delta_{
\rm bulk}\big(1-\frac{\sin(2\kf x)}{2\kf x}\big)$ as
expected for the case of pairing concentrated near the Fermi surface.  The
pairing shown in the main panel and inset 
both agree well with this approximate formula and with the semi-analytic 
approach of the preceding subsection, as
we already mentioned above. Similar behavior
was found in the work of Stojkovi\'c and Valls, who studied superconductor-insulator interfaces at low-$T_{\rm c}$ (see Fig.~2 of Ref.~\onlinecite{StojkovicPRB1993}).

Remarkably, although
Fig.~\ref{fig:LowTc Delta} is based on a full
nonlinear calculation, it looks identical
to the BCS-regime pairing kernel eigenstates 
at low $q$, as shown in, e.g., the solid blue curve in 
Fig.~\ref{fig:LowTcEigenstates}).  Thus, although
the pairing kernel eigenstates all oscillate with
wavevector $q$, if we \lq\lq zoom-in\rq\rq\ close
enough, they look like the nonlinear solution
in Fig.~\ref{fig:LowTc Delta} (which is constant
at large $x$).  
This tells us that, well below $\tc$ in the BCS regime, the local edge pairing indeed reflects the edge behavior of the eigenfunctions.

Figure~\ref{fig:TemperatureEvolutionAtLowTc}   shows the
temperature evolution as $T\to \tc$ for the same coupling value (or equivalently, the same $\tc$) as in Fig.~\ref{fig:LowTc Delta}.  Here, the
blue curve has $T$ approximately
the same as in Fig.~\ref{fig:LowTc Delta},
with the red dashed and green
dot-dashed curves showing progressively increasing 
$T$.   We see that as $T$ approaches  $\tc$, the pairing
varies over a much longer temperature-dependent length scale, roughly 
consistent with the expectations of
Ginzburg-Landau theory (although
the curves still all show a sharp
drop in pairing near the edge).

\begin{figure}[t]
  \centering
  \includegraphics[width=\columnwidth]{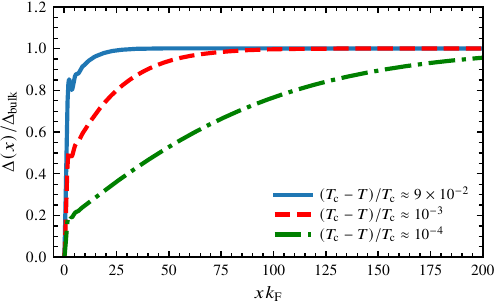}
  \caption{(Color online) The three curves show the 
  temperature evolution of $\Delta(x)$ as $T \to T_{\rm c}$ from below, at a relatively low $T_{\rm c }$ ($\beta_{\rm{c}}\mu =10$), with the blue solid
  curve showing the lowest $T$, the red dashed higher and the green dot-dashed the highest.  Thus, as $T \to T_{\rm c}$ the pairing near the boundary is increasingly suppressed.  All curves show a rapid onset of pairing near $x=0$ over a length scale  
  $\sim (2k_{\rm F})^{-1}$, along with a slow relaxation to the bulk value on the order of the coherence length.  The rapid onset at the boundary is indicative of a sharp Fermi surface at this temperature. Figure \ref{fig:TemperatureEvolutionNearUnitarity} shows the same temperature evolution for a much higher $T_{\rm c}$, where such Fermi surface features are almost entirely washed out. 
  }
\label{fig:TemperatureEvolutionAtLowishTc}
\end{figure}

Since $\tc$ is so low for 
Figs.~\ref{fig:LowTc Delta} and 
\ref{fig:TemperatureEvolutionAtLowTc},
it is numerically challenging to
get very close to $\tc$.  In 
Fig.~\ref{fig:TemperatureEvolutionAtLowishTc}, we show the temperature
dependence of the pairing vs. 
position curves at a somewhat 
stronger coupling, still in the
BCS regime but with $\kb \tc/\mu \sim 0.1$.  As in the previous, 
the lowest $T$ curve (solid blue) shows nearly uniform bulk pairing 
with a sharp drop to zero on 
a scale  $\sim 1/\kf$, similar to Fig.~\ref{fig:LowTc Delta},
although with suppressed
Friedel oscillations.  With increasing $T\to \tc$,
as shown in the red dashed and green dot-dashed curves, we again see
a temperature-dependent suppression of pairing consistent with the 
expectation of GL theory.

\begin{figure}[t]
  \centering
  \includegraphics[width=\columnwidth]{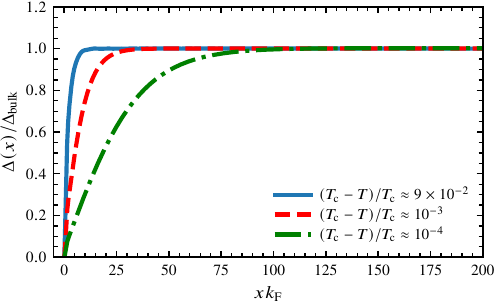}
  \caption{(Color online) Shown here is the evolution, as $T \to T_{\rm c}$ from below, of
  $\Delta(x)$ for the case of a system close to the unitary point, with $\beta_{\rm{c}}\mu =2$. Similarly to Fig.~\ref{fig:TemperatureEvolutionAtLowishTc},  the pairing near the boundary is suppressed.   Although at this high temperature the Fermi surface has almost been almost entirely washed away, and thus the pairing raises very smoothly from the boundary to its bulk value.  
  In this stronger-coupling regime the pairing curves 
  more closely follow the predictions of GL theory, as
  also shown in 
  Fig.~\ref{fig: Comparison to GL at High Tc}.
 }
\label{fig:TemperatureEvolutionNearUnitarity}
\end{figure}

Next we turn to the strong coupling regime near unitarity, 
in which the transition temperature is much higher.  Figure~\ref{fig:HighTc Delta} shows the pairing vs position
for the case of $\kb \tc/\mu  = 0.5$ and the temperature somewhat
below $\tc$ (i.e., $\tc-T\simeq 0.09 \tc$).  Here,  the dashed blue
and solid red curves show the semi-analytic and full numerical
curves, which show reasonable agreement.  Interestingly, we 
see no significant Friedel oscillations, with a much smoother
edge behavior (in comparison to
the low-$\tc$ curves). 

\begin{figure}[t]
  \centering
  \includegraphics[width=\columnwidth]{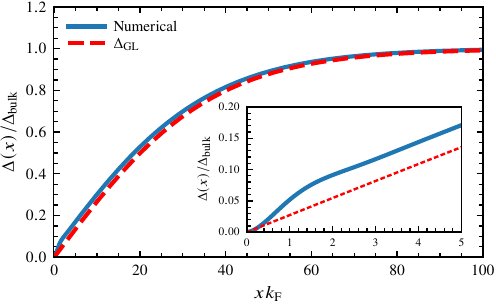}
  \caption{(Color online) The solid line shows  the  local pairing  in the stronger coupling regime ($\beta_{\rm{c}}\mu =2$) and at a temperature very close to $T_{\rm c}$; $(T_{\rm c}-T)/T_{\rm c}\approx 10^{-4}.$ For comparison the Ginzburg–Landau (GL) solution $\Delta_{\rm GL}(x)=\Delta_{\rm bulk}\tanh(x/(\sqrt{2}\xi)$ is also shown, where $\xi$ is the coherence length at temperature $T$, see Appendix \ref{app:GL} for details of GL theory. The inset shows behavior of pairing for both near the boundary showing a slight difference.  The GL profile goes linearly to zero, while the full numerical solution approaches zero quadratically, but then turns over to a linear increase after a distance of approximately $(2k_{\rm F})^{-1}$.      }
  \label{fig: Comparison to GL at High Tc}
\end{figure}

Fig.~\ref{fig:TemperatureEvolutionNearUnitarity} shows the temperature
evolution as $T\to \tc$ in the stronger coupling region, 
with increasing $T$ shown 
progressively in the solid blue, 
red dashed, and green dot-dashed 
curves. 
  In contrast
to the lower-$\tc$ curves that
exhibited two length scales over
which the pairing changed, 
these curves show no significant
edge effects. In  
Fig.~\ref{fig: Comparison to GL at High Tc}, we  directly compare
a curve in this strongly-interacting
regime to the prediction of
Ginzburg-Landau theory, showing 
close agreement.  

\section{Concluding Remarks}
\label{sec:concl}
In this paper we have studied superfluid pairing 
of attractively-interacting fermions 
in a semi-infinite system with one \lq\lq wall\rq\rq\ at $x=0$
as a model of edge pairing in a confined atomic superfluid
or electronic superconductor.  

Our approach treated 
the pairing instability as an eigenvalue problem, finding
continuum eigenstates of the pairing kernel at $\tc$.  In
the weak-coupling BCS limit, we
find these eigenstates exhibit strong Friedel oscillations 
at a wavevector scale $2\kf$ (consistent with the well-known
picture that BCS pairing is a Fermi-surface phenomenon).  
With increasing coupling strength towards the BEC regime, these
oscillations are suppressed.  Our
analysis also found no bound states of the 
pairing kernel.  If their energies were
below the continuum, such a bound state would
indicated a ``surface pairing'' instability 
preceding the bulk transition. 

Thus, for our model we do not find any evidence of a such a surface state, which would represent a localized  superfluid/superconductor order parameter near the boundary with a $\tc$ higher than the bulk.   
We note that other recent work has found
surface states in other related models.  For example,    Refs.~\cite{SamoilenkaBabaev2020,Hainzl2022} recently showed that the 1D version of 
the $\tc$-equation, Eq.~\eqref{Eq:kappamatrix}, produces  such a surface state.  Interestingly for contact interactions in higher dimensions,  both 2D \cite{PattonUnpublished} and 3D (this work), we find no evidence for such surface states. Furthermore, we find  that with the inclusion of the mean-field Hartree potentials, the surface states in 1D are no longer present \cite{PattonUnpublished}.

To study edge pairing below $\tc$, we generalized our model to
include a Ginzburg-Landau (GL) type nonlinearity that stabilized
pairing at large $x$, finding that the weak-coupling Friedel
oscillations persist in the region below $\tc$.

We now propose some natural extensions of our work.  Firstly,
our treatment of pairing below $\tc$ used an approximate GL 
approach. It would be interesting to extend this to a
self-consistent Bogoliubov-de Gennes (BdG) approach to pairing 
in a half-slab Fermi superfluid. Presumably, the local BCS coherence factors would also be cut-off independent, similar in nature to  how we have shown that the eigenfunctions of the $T_{\rm c}$-kernel are, while the BdG energies would remain cut-off dependent.    Additionally, while
the half-slab assumption simplified some aspects of the
calculation, it also created difficulties due to the 
continuum nature of pairing kernel eigenfunctions.  This motivates
studying edge pairing in the BEC-BCS crossover in a truly
finite box geometry.  One key difference is that we expect the box $\tc$ to be different from the bulk $\tc$ (in contrast to the 
half slab case where they are equal), 
complicating the regularization procedure 
in the box case.  
We leave these problems  for future work.

\begin{acknowledgments}
 We gratefully acknowledge discussions with Anshuman Bhardwaj.  DES acknowledges support from the National Science Foundation under Grant PHY-2208036.  
 This work was performed in part at Aspen Center for Physics, which is supported by National Science Foundation grant PHY-2210452.
\end{acknowledgments}

\appendix 
\section{Double Principal Value integration}
\label{app:dvpi}
The fact that the integral of a product of principal value distributions
yields a delta-function distribution is well-known and connected
to the Poincar\'e-Bertrand theorem~\cite{NewtonShtokhamer,Davies1990}.
Here, we present a simple demonstration of this fact using
our basis functions $\chi_Q(x)\equiv \langle x|Q\rangle$ and $\psi_k(x) \equiv \langle x|k\rangle$ 
defined above in Eq.~(\ref{Eq:cossinbasis}).  (We use
the Dirac notation versions below.)
We quote here the relevant distributional orthonormality 
relations:
\bse
\bea
\label{eq:kk}
&&\int_0^\infty {\rm d}x \psi_k(x) \psi_{k'}(x) = \pi \delta(k-k'),
\\
&& \int_0^\infty {\rm d}x \chi_Q(x) \psi_{Q'}(x) = \pi \delta(Q-Q'),
\eea
\label{eqs:kkqq}
\ese
and completeness relations:
\bse
\bea
&&\int_0^\infty \frac{{\rm d}k}{\pi} \psi_k(x) \psi_{k}(x') = \delta(x-x'),
\\
&& \int_0^\infty \frac{{\rm d}Q}{\pi} \chi_Q(x) \psi_{Q}(x') =  \delta(x-x').
\label{eq:qqcomp}
\eea
\label{eqs:kkqqcomp}
\ese
To demonstrate the necessary identity, we start with
the inner product between these basis sets, which is:
\bea
\langle k|Q\rangle  &=& \langle Q|k\rangle 
=2\int_0^\infty {\rm d}x\, \sin(kx) \cos(Qx) ,
\nonumber 
\\
&=& \curP \frac{2k}{k^2-Q^2},
\label{Eq:innerProductcossin}
\eea
with $\curP$ indicating the principal value.  One way to 
demonstrate this is to regularize the integral by including a
factor  ${\rm e}^{-\epsilon x}$, subsequently taking the limit
of $\epsilon\to 0$.
Since the sine-basis functions are delta-function normalized, their inner product is $\langle k|k'\rangle = \pi\delta(k-k')$
(see Eq.~(\ref{eq:kk})).
Inserting the identity of cosine-basis
functions , $\bone = \int_0^\infty \frac{{\rm d}Q}{\pi}|Q\rangle\langle Q|$ (which is consistent with
Eq.~(\ref{eq:qqcomp}))
we get 
\bea
\pi\delta(k-k') &=& \langle k | k'\rangle  = \int_0^\infty \frac{{\rm d}Q}{\pi} \langle k|Q\rangle
\langle Q|k'\rangle ,\nonumber 
\\
&=&
\int_0^\infty \frac{{\rm d}Q}{\pi} \curP \frac{2k}{k^2-Q^2}
\curP \frac{2k'}{(k')^2-Q^2} .
\label{integralproductprincipal}
\eea
This demonstrates that the coinciding principal value 
singularities give a delta-function result when integrated.  Since the integral in
Eq.~(\ref{integralproductprincipal}) has the same behavior near its poles as the 
one in the main text, this is sufficient to demonstrate the steps leading to Eq.~(\ref{Eq:deltainnerQ}).

\section{Ginzburg-Landau theory}
\label{app:GL}
In this section we recall Ginzburg-Landau (GL) theory
for a half slab system.  We have the GL equation 
for the pairing amplitude $\Delta(x)$ as a function of
position:
\be
 \label{eq:deltavsx}
0 = \left(-\frac{1}{2}\rho \partial_x^2  +\alpha 
\right)\Delta(x) + \beta |\Delta(x)|^2 \Delta(x),
 \ee
where $\rho>0$ and $\beta>0$ and we assume the boundary condition
$\Delta(0) = 0$.  For $T<\tc$, the parameter 
$\alpha<0$, vanishing for $T\to \tc$.  Henceforth we 
assume $\alpha<0$. 

The standard GL equation is an approximation for $\Delta(x)$ near $T_c$, under the assumption that $\Delta(x)$ is  slowly varying. In the bulk, this slow variation amounts to Taylor expanding $\curE^\Lambda(Q)$, Eq.~\eqref{eq:epsilonLambdaQ}, around $Q =0$. This relates the GL parameters $\rho$ and $\alpha$ to the microscopic description  via  $\rho = -g\partial^2_Q\curE^\Lambda(Q)\big|_{Q\to 0}$ and $\alpha = 1 - g\curE^\Lambda(0)$, where $g$ is the coupling that defines $T_{\rm c}$ and $\curE^\Lambda(0)$ is evaluated at $T<T_{\rm c}$.  One can define the GL coherence length $\xi$ as $\xi = \sqrt{\rho/(2|\alpha|)}$ and  show that it is effectively cutoff independent.

The pairing kernel for this present GL problem is simply the operator in parentheses, 
$\curK_{\rm GL} \equiv -\frac{1}{2}\rho \partial_x^2  -|\alpha|$, with eigenfunctions 
$\psi_k(x) = \sqrt{2}\sin(kx)$ and eigenvalues $E(k) = \frac{1}{2}\rho k^2 -|\alpha|$.
Note that although
the vanishing of the lowest eigenvalue correctly tells us the phase transition
$(\alpha = 0)$, we cannot easily get the shape of the pairing below $\tc$ using these eigenfunctions.  In particular, the eigenfunctions vanish in the limit $k\to 0$ where
we reach the lowest eigenvalue, just as in the full case studied in the main text. 

The solution for all $x$ is, of course, well known:
 \be
 \label{eq:tanhSolution}
\Delta(x) = \Delta_0 \tanh\left(\frac{x}{\sqrt{2}\xi}\right), 
 \ee
 where $\Delta_0 =  \sqrt{\frac{|\alpha|}{\beta}}$.
How can we use the exact eigenfunctions of the kernel to get the approximate edge behavior?
One approximate strategy is to assume  $\Delta(x) = \Delta_0 +\Delta_1(x)$, i.e., a
sum of a bulk piece $\Delta_0$ plus a correction.  Then, we plug this 
into Eq.~(\ref{eq:deltavsx}), 
Taylor expand all terms to linear order in $\Delta_1(x)$, and re-express the resulting 
equation in terms of $\Delta(x)$.  The result is:
 \be
 \label{eq:taylord1p}
 0 = -\frac{1}{2} \partial_x^2 \Delta(x) +\frac{1}{\xi^2}
 \Big(\Delta(x) - \Delta_0\Big),
  \ee
Note the steps leading to Eq.~(\ref{eq:taylord1p}) are very approximate, since we formally assumed $\Delta_1\ll \Delta_0$,
but we need $\Delta_1(0) = -\Delta_0$ to satisfy the boundary condition.  However, since 
it is linear in $\Delta(x)$, we can easily solve it via Fourier transform.  To do this, we
can express the final \lq\lq constant\rq\rq\ term of Eq.~(\ref{eq:taylord1p}) as a Fourier series
in the basis of our eigenfunctions, using 
  \be
1  = \int_0^\infty \frac{{\rm d}k}{\pi} \psi_k(x) \frac{\sqrt{2}}{k} .
  \ee
Using this in Eq.~(\ref{eq:taylord1p}), and taking the sine Fourier transform of all other terms
using $\Delta(x) = \int_0^\infty \frac{dk}{\pi}\psi_k(x) \Delta(k)$, we get the following solution
for $\Delta(k)$:
  \be
 \Delta(k)=
 \Delta_0 \frac{\sqrt{2}/k}{1+k^2\xi^2/2},
  \ee
which we note goes as $1/k$ for small $k$ (so that $\Delta(x)$ goes to a constant at large $x$, as expected).
Transforming back to real space, we get the final result
\be
\label{Eq:deltavxEXP}
\Delta(x) = \Delta_0\Big(1- {\rm e}^{-\sqrt{2}x/\xi}\Big),
\ee
 that is qualitatively similar to the exact result Eq.~(\ref{eq:tanhSolution}).  This gives a
 strategy for obtaining the approximate edge behavior of the pairing amplitude which we mirror in 
 Sec.~\ref{sec:sals}.

      \section{Bound
      on the spectrum of $\curK^\Lambda$}
    \label{app:bound} 
      In this section we
      derive a bound on the spectrum of the pairing kernel $\curK^\Lambda$. 
      We assume the presence of a normalized bound state $|\psi\rangle$.
      The eigenvalue of such a purported bound state would be
      $E=\langle \psi | \curK^\Lambda |\psi\rangle$.  
      Then, using Eq.~(\ref{Eq:kappamatrix}) for the kernel in the cosine basis,  we get:
      \bea
   &&   
\hspace{-1cm}
E=    \int_0^\infty \frac{{\rm d}Q}{\pi}\int_0^\infty \frac{{\rm d}Q'}{\pi}
      \frac{1}{2}f_{Q,Q'}^\Lambda \Big[
        \psi(Q)^2 - \psi(Q)\psi(Q')\Big],
      \eea
      where $\psi(Q) = \langle Q|\psi\rangle$ and we used Eq.~(\ref{eq:epsilonLambdaQ}) to simplify the expression.  Since
      $f_{Q,Q'}^\Lambda$ is symmetric under interchanging $Q$ and $Q'$,
      we can replace the first term in square brackets by
      $\psi(Q)^2 \to \frac{1}{2}\big( \psi(Q)^2+\psi(Q')^2\big)$.
      Now we have:
      \bea
     E=  \int_0^\infty \frac{{\rm d}Q}{\pi}\int_0^\infty \frac{{\rm d}Q'}{\pi}
      \frac{1}{4}f_{Q,Q'}^\Lambda \Big(
      \psi(Q) -\psi(Q')\Big)^2,
      \label{eq:needintegrand}
        \eea
       Next, we use  
       the Cauchy-Schwarz inequality
($|\langle a|b\rangle|^2 \leq \langle a|a\rangle
        \langle b |b\rangle$)
       to place a bound on the integrand of
       Eq.~(\ref{eq:needintegrand}). 
If we write $a$ and $b$ as $n$-component real vectors, 
        this can be written as:
        \be
        \left(\sum_{i = 1}^n a_i b_i \right)^2 \leq \left(\sum_{i=1}^n a_i^2\right)
        \left(\sum_{i=1}^n b_i^2\right).
        \ee
        Applying this to the case of $n=2$ with $a_1 = \psi(Q)$,
        $a_2 =-\psi(Q')$, and $b_i=1$, we obtain:
        \bea
        \Big( \psi(Q) -\psi(Q')\Big)^2\leq 2\Big(\psi(Q)^2+\psi(Q')^2\Big).
        \label{eq:boundwith2}
        \eea
        Since $f_{Q,Q'}^\Lambda<0$, which follows since the integrand of Eq.~(\ref{eq:integrandF}) is negative, we have:
         \be
         f_{Q,Q'}^\Lambda\Big( \psi(Q) -\psi(Q')\Big)^2\geq
         2 f_{Q,Q'}^\Lambda\Big(\psi(Q)^2+\psi(Q')^2\Big),
        \ee
        so that we have 
         \bea
        E \geq \int_0^\infty \frac{{\rm d}Q}{\pi}\int_0^\infty \frac{{\rm d}Q'}{\pi}
         \frac{1}{2}f_{Q,Q'}^\Lambda \Big(\psi^2(Q) + \psi^2(Q')\Big).
         \eea
         For each of the two terms on the right side, we can evaluate the
         integrals to get
         \be
      E   \geq 2
         \int_0^\infty
         \frac{{\rm d}Q}{\pi}\curE^\Lambda(Q) \psi^2(Q) .
         \ee
         Finally using that $\curE^\Lambda(0)\leq\curE^\Lambda(Q)$, and that
         $\psi(Q)$ is
         assumed normalized, so that $\int_0^\infty \frac{{\rm d}Q}{\pi} \psi^2(Q) =
         1$, we get
         the final bound 
         \be
\langle \psi|\curK^\Lambda |\psi\rangle \geq 2\curE^\Lambda(0).
         \ee
         Since the continuum states of $\curK^\Lambda$ have energy bounded by
             $\curE^\Lambda(0)$ (note 
             $\curE^\Lambda(0)<0$), with this we
         have proven that any bound states below the continuum (which would represent a \lq\lq surface\rq\rq\ pairing instability preceding the bulk pairing instability)
         must have energy
         $ 2\curE^\Lambda(0)\leq E_{\rm bound}\leq\curE^\Lambda(0)$.

\section{Numerical details}
\label{app:numdet}
In this Appendix we describe details of our
numerical approach to the pairing kernel
eigenvalue problem and the nonlinear pairing 
problem.  

We start by noting that
the continuous Fourier-cosine transform pair is defined as
\begin{equation}
\label{Continous Cosine Transform x to k}
y(k) = A \int\limits_0^{\infty}{\rm d}x\, y(x)\cos(kx),
\end{equation}
and
\begin{equation}
\label{Continous Cosine Transform k to x}
y(x) = B \int\limits_0^{\infty}{\rm d}k\, y(k)\cos(kx),
\end{equation}
where the constants $A$ and $B$ are chosen to satisfy the inversion requirement;  $AB = 2/\pi$. There are several conventional choices for this. For numerical  convenience, in the following we set $A = B = \sqrt{2/\pi}$. Note, this is a different convention than the main text. 

To  numerically solve the $T_{\rm c}$-gap equation in momentum space and then  transform back to real space requires one to discretize  the  cosine transform pair and the $T_{\rm c}$-integral equation itself. As the solutions of the gap equation at $T_{\rm c}$ are just the eigenvectors of the kernel ${\cal K}^{\Lambda}$, to numerically preserve the orthonormality of the eigenvectors in both real and momentum space, we use the following quadrature for the discrete cosine transform (DCT).
\subsection{Discrete cosine transform}
\label{DCT definition}
Let $L$ be the effective finite system size cutoff in  real space. This should be chosen to be much larger than all other length scales, e.g., the Fermi wavelength and the coherence length. Let $N$ be the number of discrete sample points in both the physical space and momentum space. The grid size in real space is then defined as $\Delta x = L/(N-1)$ and in momentum space as $\Delta k = \pi/L$. The sample points are then given by $x_n = n\Delta x$ and $k_n= n\Delta k$ with $n = 0,1,\ldots, N-1$. This introduces an upper cutoff in momentum space given by $k_{\rm max} = (N-1)\pi/L$, which should be much larger than $k_{\rm F}$. For all results shown here, we have used an $L$ ranging from $10^{3} k^{-1}_{\rm F}$ to $2\times10^4 k^{-1}_{\rm F}$ and  $N$ ranging from $10^4$ to $ 10^5$.

The continuous  Fourier-cosine transform pair, Eqs.~\eqref{Continous Cosine Transform x to k} and \eqref{Continous Cosine Transform k to x}, is then approximately given by 
\begin{equation}
y(k^{}_{n}) \approx \sqrt{\frac{2}{\pi}}\sum_{m = 0}^{N-1} y(x^{}_{m})\cos(k^{}_{n}x^{}_{m})w^{x}_{m},
\end{equation}
and
\begin{equation}
y(x^{}_{n}) \approx \sqrt{\frac{2}{\pi}}\sum_{m = 0}^{N-1} y(k^{}_{m})\cos(k^{}_{n}x^{}_{m})w^{k}_{m},
\end{equation}
where the weight functions $w_n$ are defined as
\begin{equation}
\label{DCT weights x}
w^{x}_{n} = 
\begin{cases}
\Delta x/2 & \text{if } n = 0 \text{ or } n = N-1  \\\\
\Delta x & \text{if } n =  1,\ldots, N-2.
\end{cases}
\end{equation}
and similarly  
\begin{equation}
\label{DCT weights k}
w^{k}_{n} = 
\begin{cases}
\Delta k/2 & \text{if } n = 0 \text{ or } n = N-1  \\\\
\Delta k & \text{if } n =  1,\ldots, N-2.
\end{cases}
\end{equation} 
The inner product $\langle f|g \rangle$ between two functions in this discrete space is defined with respect to the weight functions $w^{}_{n}$, Eqs.~\eqref{DCT weights x}-\eqref{DCT weights k}, i.e., 
\begin{equation}
\label{inner product in k and x space}
\langle f|g \rangle = \sum_{n = 0}^{N-1} f(x^{}_{n})g(x^{}_{n})w^{x}_{n} = \sum_{n = 0}^{N-1} f(k^{}_{n})g(k^{}_{n})w^{k}_{n},
\end{equation}
where we have used the following identity to arrive at the second equality
\begin{equation}
\sum_{m=0}^{N-1} \cos(k^{}_{n}x^{}_{m})\cos(k^{}_{n'}x^{}_{m})w^{x}_{m} = \frac{\pi}{2}\frac{\delta^{}_{n,n'}}{w^{k}_{n}}.
\end{equation}
\subsection{At $T_{\rm c}$}
At $T_{\rm c}$, the gap equation is an integral eigenvalue problem given in $k$-space as (Again, note we have slightly different $\pi$ conventions relative to the main text)
\begin{equation}
\frac{1}{\lambda}\Delta(k)= \int\limits_{0}^{\infty}{\rm d}k'\, {\cal K}^{\Lambda}(k,k')\Delta(k'),	
\end{equation} 
where
\begin{equation}
	{\cal K}^{\Lambda}(k,k') ={\cal E}^{\Lambda}(k)\delta(k-k') -\frac{1}{2\pi}f^{\Lambda}_{k,k'}.
\end{equation}
Using the definitions of the DCT from Sec.~\ref{DCT definition} of the Appendix the discretized  $T_{\rm c}$-equation becomes 
\begin{equation}
	\frac{1}{\lambda}\Delta(k^{}_{n}) = \sum_{m =0}^{N-1}{\cal K}^{\Lambda}_{}(k^{}_{n},k^{}_{m})w^{k}_{m}\Delta(k^{}_{m}),
\end{equation}
where
\begin{equation}
	{\cal K}^{\Lambda}(k^{}_{n},k^{}_{m})={\cal E}^{\Lambda}_{}(k^{}_{n})\frac{\delta^{}_{n,m}}{w^{k}_{m}} -\frac{1}{2\pi}f^{\Lambda}_{k^{}_{n},k^{}_{m}}.
\end{equation}
Because of the weight function $w^{k}_{m}$, the matrix of the eigenvalue problem, ${\cal K}^{\Lambda}_{}(k^{}_{n},k^{}_{m})w^{k}_{m}$, is no longer symmetric.  We can turn this into a Hermitian matrix eigenvalue problem by introducing an auxiliary function $u(k)$, which is related to the pairing function via $\Delta(k^{}_{m}) = u(k^{}_{m})/\sqrt{w^{k}_{m}}$. Expressing the eigenvalue equation in terms of this auxiliary function leads to the following  eigenvalue problem 
\begin{equation}
	\frac{1}{\lambda}u(k^{}_{n}) = \sum_{m =0}^{N-1}{\cal K}^{\Lambda}_{}(k^{}_{n},k^{}_{m})\sqrt{w^{k}_{n}w^{k}_{m}}u(k^{}_{m}),
\end{equation}
now with a symmetric matrix, ${\cal K}^{\Lambda}_{}(k^{}_{n},k^{}_{m})\sqrt{w^{k}_{n}w^{k}_{m}}$.
After diagonalization most numerical libraries return a set  of orthonormal eigenvector $u^{}_i(k^{}_{m})$, i.e., 
\begin{equation}
\label{orthonormal Us}
	\sum_{m=0}^{N-1}u^{}_i(k^{}_{m})u^{}_j(k^{}_{m}) = \delta^{}_{i,j},
\end{equation}
along with their associated eigenvalues $1/\lambda_i$, which we find to numerical accuracy  are given by $1/\lambda_i \approx {\cal E}^{\Lambda}(q^{}_{i})$. The minimum eigenvalue defines the critical coupling $g_{\rm c}$ for a chosen $T_{\rm c}$.
Then, we obtain our eigenfunctions from:
\begin{equation}
	\Delta^{}_{q^{}_i}(k^{}_{m})= \frac{u^{}_i(k^{}_{m})}{\sqrt{w^{k}_{m}}}.
\end{equation}
Using Eqs.~\eqref{orthonormal Us} and \eqref{inner product in k and x space}, one can produce a set of orthonormal eigenfunctions in both real and $k$-space, i.e.,
\begin{equation}
\label{Numerical orthonormal system}
	\sum_{m=0}^{N-1}\Delta^{}_{q^{}_i}(k^{}_{m})\Delta^{}_{q^{}_j}(k^{}_{m})w^{k}_{m} = \sum_{m=0}^{N-1}\Delta^{}_{q^{}_i}(x^{}_{m})\Delta^{}_{q^{}_j}(x^{}_{m})w^{x}_{m}= \delta^{}_{i,j},
\end{equation}
where 
\begin{equation}
	\Delta^{}_{q^{}_i}(x^{}_{n}) = \sqrt{\frac{2}{\pi}}\sum_{m = 0}^{N-1} \Delta^{}_{q^{}_i}(k^{}_{m})\cos(k^{}_{m}x^{}_{n})w^{k}_{m}.
\end{equation}

This step is crucial for going below  $T_{\rm c}$, as we  solve the nonlinear problem that occurs below  $T_{\rm c}$ by expanding  the local pairing $\Delta(x)$ in the eigenbasis defined by $\Delta_{q^{}_i}(x^{}_{n})$, and then solve the resulting nonlinear equation for the expansion coefficients. 
\subsection{Below $T_{\rm c}$}
To analyze the regime of $T<\tc$, 
we extend the linear $T_{\rm c}$-equation by adding a Ginzburg-Landau-like (GL) nonlinear term to stabilize the bulk pairing far away from the boundary. Thus we seek to solve the following nonlinear integral equation
\begin{equation}
\label{Below Tc nonlinear equation}
 \Delta(x)= g^{\Lambda}_{\rm c}\int\limits_{0}^{\infty}{\rm d}x'\, {\cal K}^{\Lambda}(x,x')\Delta(x')+g^{}_{\rm c}\beta^{}_{\rm GL}|\Delta(x)|^{3},   
\end{equation}
where $1/g^{\Lambda}_{\rm c} ={\cal E}^{\Lambda}(q^{}_0)$ is the cutoff-dependent critical coupling at $T_{\rm c}$ and the parameter $\beta_{\rm GL}$, the
coefficient of the nonlinear term, 
controls the bulk value of 
$\Delta(x)$.  Within standard Ginzburg-Landau theory
$\beta^{}_{\rm GL} = m k_{\rm F} 7\zeta(3) \beta^2/(16 \pi^4)$, where $\zeta(n)$ is the Riemann zeta function and $\beta = 1/(k_{\rm B}T)$, but here $\beta^{}_{\rm GL}$  remains an arbitrary parameter, kept for dimensional reasons. 

We solve this by expanding the local pairing into the space spanned by the eigenstates of the   $T_{\rm c}$-equation, i.e., the basis defined by $\Delta_{q^{}_i}(x^{}_{n})$.  In practice we only work in a subspace spanned by the lowest $N_{\rm nl}$ eigenstates, where $N_{\rm nl}\ll N$. For all results shown here we used the lowest 200 to 500 states.  The local pairing is then approximately given by 
\begin{equation}
\label{Eigenbasis expansion for Delta}
    \Delta(x_n) \approx \sum_{i = 0}^{N_{\rm nl}}c_{q^{}_i}\Delta_{q^{}_i}(x^{}_{n}),
\end{equation}
where $c_{q^{}_i}$ are the unknown expansion coefficients. 

Inserting Eq.~\eqref{Eigenbasis expansion for Delta} into \eqref{Below Tc nonlinear equation}, and  using 
\begin{equation}
    \int\limits_{0}^{\infty}{\rm d}x'\, {\cal K}^{\Lambda}(x,x')\Delta_{q}(x') \approx {\cal E}^{\Lambda}(q)\Delta_{q^{}}(x^{}),
\end{equation}
where now ${\cal E}^{\Lambda}(q)$ is now evaluated at $T$ instead of $T_{\rm c}$, we arrive at 
\begin{equation}
    \sum_{i = 0}^{N_{\rm nl}}c^{}_{q^{}_i}\Delta_{q^{}_i}(x^{}_{n}) = g^{\Lambda}_{\rm c}\sum_{i = 0}^{N_{\rm nl}}{\cal E}^{\Lambda}(q^{}_i)c_{q^{}_i}\Delta_{q^{}_i}(x^{}_{n})+g^{\Lambda}_{\rm c}\beta^{}_{\rm GL}|\Delta(x^{}_{n})|^{3}.
\end{equation}
Multiplying  through by $\Delta_{q^{}_j}(x^{}_{n})$ and integrating using the quadrature rules and orthonormality given by Eq.~\eqref{inner product in k and x space} and Eq.~\eqref{Numerical orthonormal system} leads to a system of nonlinear equations for the unknown expansion coefficients, $c_{q^{}_j}$,  
\begin{equation}
    c_{q^{}_j} = g^{\Lambda}_{\rm c}{\cal E}^{\Lambda}(q^{}_j)c_{q^{}_j}+g^{\Lambda}_{\rm c}\beta^{}_{\rm GL}\sum_{n = 0}^{N-1} \Delta_{q^{}_j}(x^{}_{n})|\Delta(x^{}_{n})|^{3}w^{x}_{n}.
\end{equation}
As described in the body of the article one can replace the critical coupling and spectrum with their renormalized values: $g^{\Lambda}_{\rm c}\to g^{}_{\rm c}$ and ${\cal E}^{\Lambda}(q)\to {\cal E}^{R}(q)$. Doing so finally leads to 
\begin{equation}
    c_{q^{}_j} = g^{}_{\rm c}{\cal E}^{R}(q^{}_j)c_{q^{}_j}+g^{}_{\rm c}\beta^{}_{\rm GL}\sum_{n = 0}^{N-1} \Delta_{q^{}_j}(x^{}_{n})|\Delta(x^{}_{n})|^{3}w^{x}_{n}.
\end{equation}
We solve this system of equations by casting it into a nonlinear root finding problem for the $N_{\rm nl}$ coefficients.

\begin{figure}[t]
  \centering
  \includegraphics[width=\columnwidth]{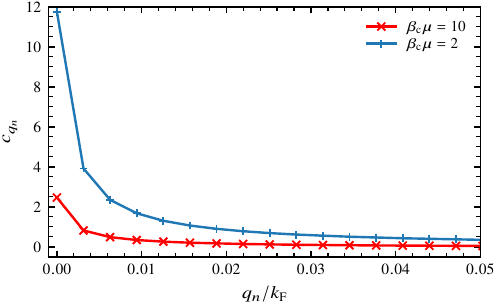}
  \caption{(Color online) The numerically obtained coefficients $c_{q_n}$ of the eigenfunction expansion of $\Delta(x)$ for $T< T_{\rm c}$, Eq.~\eqref{Eigenbasis expansion for Delta}, are shown,   for a couple of representative critical  temperatures  $\beta_{\rm c}\mu = 10$ and $2$ with temperature $\beta\mu = 1.1\beta_{\rm c}\mu$. As can be seen, only the lowest few eigenstates significantly contribute to the pairing below $T_c$, with the ground state $c_{q}$ being the most dominate.}  
  \label{fig:Eigen expand coeffs}
\end{figure}

Figure \ref{fig:Eigen expand coeffs} shows the numerical values obtained for the expansion coefficients, $c_{q^{}_j}$,
for two different, relatively high, transition temperatures. As one can see, even at such large transition temperatures only the lowest few eigenstates significantly contribute to the local pairing, with the ground state, $c_0$, being the most dominant. The contribution of the higher states rapidly goes to zero. As $T_{\rm c}$ is lowered this suppression of the contribution from higher states becomes even more pronounced. 

\end{document}